\documentclass[12pt]{iopart}
\usepackage{iopams}
\usepackage{harvard}
\usepackage{graphicx}
\usepackage{booktabs}
\newcommand{\dgr}{$^{\circ}$}

\begin{document}

\title{Benchmark Calculations of Electron Impact Electronic Excitation of the Hydrogen Molecule}

\date{\today}

\author{
T. Meltzer$^1$,
J. Tennyson$^1$,
Z. Ma{\v{s}}{\'{\i}}n$^2$,
M. C. Zammit$^3$,
L. H. Scarlett$^4$,
D. V. Fursa$^4$ \&
I. Bray$^4$.
}

\address{
  $^1$Department of Physics and Astronomy, University College London, London WC1E 6BT,
  United Kingdom.\\
  $^2$Institute of Theoretical Physics, Faculty of Mathematics and Physics, Charles University, V
  Hole\v{s}ovi\v{c}k\'{a}ch 2, 180 00 Prague 8, Czech Republic.\\
  $^3$Theoretical Division, Los Alamos National Laboratory, Los Alamos, NM 87545, USA.\\
  $^4$Curtin Institute for Computation and Department of Physics and Astronomy, Curtin University,
  Perth, WA 6102, Australia.
}

\ead{j.tennyson@ucl.ac.uk}

\begin{abstract}
 We present benchmark integrated and differential cross-sections for electron collisions with H$_2$
 using two different theoretical approaches, namely, the R-matrix and molecular convergent
 close-coupling (MCCC). This is similar to comparative studies conducted on electron-atom collisions
 for H, He and Mg. Electron impact excitation to the $b \ ^3\Sigma_u^+$, $a \ ^3\Sigma_g^+$, $B \
 ^1\Sigma_u^+$, $c \ ^3\Pi_u$, $EF \ ^1\Sigma_g^+$, $C \ ^1\Pi_u$, $e \ ^3\Sigma_u^+$, $h \
 ^3\Sigma_g^+$, $B' \ ^1\Sigma_u^+$ and $d \ ^3\Pi_u$ excited electronic states are considered.
 Calculations are presented in both the fixed nuclei and adiabatic nuclei approximations, where the
 latter is shown only for the $b \ ^3\Sigma_u^+$ state. Good agreement is found for all transitions
 presented. Where available, we compare with existing experimental and recommended data.
\end{abstract}

\maketitle

\ioptwocol

\section{Introduction}
\label{sec:introduction}

Molecular hydrogen is one of the simplest, most abundant molecules in the Universe. Understanding of
how it interacts with its surroundings is of vital importance for a large variety of physical
systems, both naturally occurring and man-made e.g., fusion plasmas, planetary atmospheres and
interstellar medium. In these environments, H$_2$ molecules are subject to frequent collisions with
low to high-energy electrons.

The equations that govern electron-molecule collisions are well understood; however, accurate and
reliable cross-sections for the different processes that can occur are few and far between. Several
recommended cross-section datasets for H$_2$ have been assembled and published
\cite{Tawara1990,Yoon2008,jt647}, and yet, in their most recent review, \citeasnoun{Anzai2012} note
that benchmark cross-sections are still not available for a variety of cases. Thus far the vast
majority of recommended H$_2$ data are based on experimental results. However, due to practical
reasons these data can not always be obtained via experiment. For example, the required target may
be unstable (short-lived), or hazardous, or both e.g., T$_2$.

Furthermore, it is often difficult to obtain \emph{complete} sets of data that contain all the
cross-sections of interest across the required energy ranges. Therefore we must often rely on theory
to provide this information. In addition, if cross-sections are required from an initial state other
than the ground state then theory is presently the only realistic option.

In this work we use molecular convergent close-coupling (MCCC) theory and R-matrix theory to produce
a set of high-accuracy, benchmark cross-sections for electron impact electronic excitation. This is
similar in spirit to the convergent close-coupling (CCC) and R-matrix comparisons for 1 and 2
(active) electron atomic systems namely H \cite{Bartschat1996}, He \cite{Lange2006} and Mg
\cite{Bartschat2010}. A similar theoretical benchmark for total cross-sections for excitation to the
$b \ ^3\Sigma_u^+$ state was performed using the Schwinger variational \cite{Lima1985}, linear
algebraic approach \cite{Schneider1985} and R-matrix \cite{jt38} approaches. It is important to note
that this benchmark was a theoretical benchmark of a two-state close-coupling calculation, and was
not intended to produce convergent cross-sections. The previous R-matrix calculation was extended by
\citeasnoun{Branchett1990a} to include the first six excited electronic states, giving an improved
integrated cross-section and subsequently differential cross-sections \cite{Branchett1991}.

Both the MCCC and R-matrix methods are well established and tested. Therefore, below we only
summarise the relevant features of each theory rather than providing a thorough derivation. For a
complete description of the MCCC and R-matrix theories, the reader is directed to previous work;
\citeasnoun{Zammit2017} and \citeasnoun{jt474} respectively.

Where data are available we compare with experiment. For example there are integrated and
differential cross-sections available for some of the lower-lying excited states at intermediate
(14 eV to 17.5 eV) \cite{Hargreaves2017} and higher energies (17.5 eV to 30 eV) \cite{Wrkich2002}.
As well as work carried out by \citeasnoun{Muse2008} which provides elastic cross-sections from 1 eV
up to 30 eV. Also, in a recent comparison between theory and experiment, \citeasnoun{Zawadzki2018}
provides cross-sections for the $X \ ^1\Sigma_g^+ \rightarrow b \ ^3\Sigma_u^+ $ transition.

\section{Method}%
\label{sec:method}

\subsection{R-Matrix}%
\label{sub:r_matrix}

For the calculations we utilise the UKRMol+ suite of codes \cite{jt772}. This new and improved
version of the former UKRMol has been successfully used for a variety of molecular targets such as
BeH \cite{Darby-Lewis2017}, CO \cite{jtCO} and pyrimidine \cite{Regeta2016}. The most notable
difference between UKRMol+ and UKRMol is the implementation of B-spline type orbital (BTO) basis
functions allowing the user to select a Gaussian type orbital (GTO) only, mixed BTO/GTO or BTO only
representation of the continuum. Use of BTOs greatly extends the range of the possible R-matrix
radius. Here we use a BTO-only continuum, a large molecular R-matrix radius of $a=100$ $a_0$ and a
triply-augmented target basis set especially designed for Rydberg-like orbitals.

\subsubsection{Target Model}
\label{ssub:target_model}

\begin{figure}[htpb]
  \centering
  \includegraphics[width=1.0\linewidth]{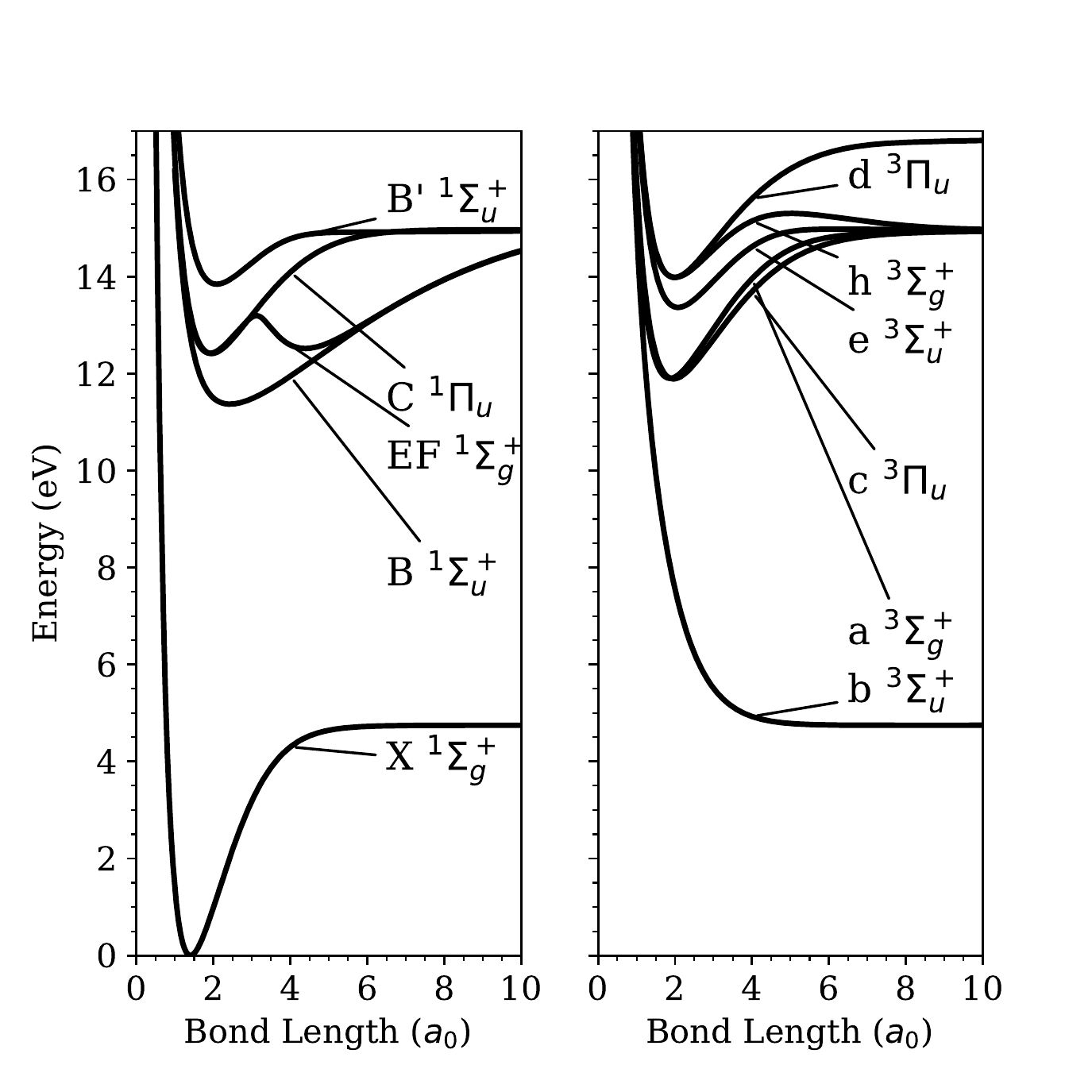}
  \caption{
   Potential energy curves for the ground state and the lower-lying excited states, relative to the
   ground state equilibrium energy. Data from \citeasnoun{Kolos1986}, \citeasnoun{Staszewska1999},
   \citeasnoun{Staszewska2002}, \citeasnoun{Wolniewicz1994} and \citeasnoun{Wolniewicz2003}.
  }
  \label{fig:PEC}
\end{figure}

The R-matrix method relies on a balanced description of the target and scattering wavefunctions, $N$
and $N+1$ respectively \cite{jt189}. Where $N$ is the number of electrons in the target. Molecular
hydrogen is a two electron system. Therefore we have aimed to use the most comprehensive models
available in each case. Full-CI is the hallmark of accuracy in electronic structure methods and it
provides an exact solution to the Schr\"{o}dinger equation within a given finite-sized one-electron
basis set. This method is used with an augmented Dunning basis set, especially designed to describe
Rydberg-type excitations in molecules, x-aug-cc-pVXZ \cite{Dunning1989,Woon1994}. x-aug signifies
that the basis set is doubly, triply, quadruply-augmented where x = d, t, q, etc. Triply augmented
means that three additional, even-tempered basis functions are added for each angular symmetry
available in the original cc-pVXZ set. Traditional Dunning basis sets, cc-pVXZ, are correlation
consistent and hence provide a systematic way of approaching the complete basis set limit as the
number X of zeta functions is increased. Preliminary work found that moving from a singly augmented
basis set to a triply augmented basis set had a more profound effect on the target description than
increasing the number of zeta functions i.e., pVXZ for X = D, T, Q, etc. For the R-matrix
calculations, presented in this work, we found that t-aug-cc-pVTZ was the optimal choice (tAVTZ
hereafter).

As mentioned previously, we are using Full-CI and the tAVTZ basis set. Therefore, our target model
(in D$_{2h}$ symmetry) can be expressed as $(19,9,9,4,19,9,9,4)^2$. Using this model we can solve
the $N$-electron problem and calculate target state energies (on average, 400 configuration state
functions are generated per molecular spin-space symmetry). Target state energies were calculated at
the equilibrium geometry $R_e=1.40 \ a_0$ to compare with accurate structure calculations of
\citeasnoun{Kolos1986}, \citeasnoun{Staszewska1999}, \citeasnoun{Staszewska2002},
\citeasnoun{Wolniewicz1994}, \citeasnoun{Wolniewicz2003} and the MCCC calculations of
\citeasnoun{Zammit2017a} (see Table \ref{tab:target_energies_Re}). Potential energy curves from the
aforementioned references are also provided in Fig. \ref{fig:PEC}.

For the excited states considered in this work, the R-matrix method produces more accurate target
states than the spherical MCCC method. This is due to the difference in how the target is expanded
in the two methods. MCCC uses single centre expansion, which performs worse for lower target states,
however it quickly improves for the higher lying, Rydberg-like states. The R-matrix method however
uses a linear combination of atom-centered GTOs. This generally performs better for the ground and
low-lying excited states and in this work it performs well for all the states listed in Table
\ref{tab:target_energies_Re}.

\begin{table}[t]
  \begin{center}
    \caption{ Absolute target energies, E (a.u.), at the equilibrium bond length $R_e=1.40 \ a_0$
    compared to accurate electronic structure calculations.}
    \label{tab:target_energies_Re}

    \vspace{5pt}

    \begin{tabular}{rccc}
      \toprule
                            & \multicolumn{3}{c}{E (a.u.)} \\
      \cline{2-4}
      state                 & Ref        & RM$^f$ & MCCC$^g$\\
      \midrule
      $X   \ ^1 \Sigma_g^+$ & -1.174$^a$ & -1.173 & -1.162 \\
      $b   \ ^3 \Sigma_u^+$ & -0.784$^b$ & -0.784 & -0.770 \\
      $a   \ ^3 \Sigma_g^+$ & -0.714$^b$ & -0.713 & -0.710 \\
      $B   \ ^1 \Sigma_u^+$ & -0.706$^c$ & -0.705 & -0.697 \\
      $c   \ ^3 \Pi_u$      & -0.707$^b$ & -0.706 & -0.701 \\
      $EF  \ ^1 \Sigma_g^+$ & -0.692$^d$ & -0.691 & -0.687 \\
      $C   \ ^1 \Pi_u$      & -0.689$^e$ & -0.688 & -0.683 \\
      $e   \ ^3 \Sigma_u^+$ & -0.644$^b$ & -0.643 & -0.640 \\
      $h   \ ^3 \Sigma_g^+$ & -0.630$^b$ & -0.630 & -0.628 \\
      $d   \ ^3 \Pi_u$      & -0.629$^b$ & -0.628 & -0.626 \\
      $B'  \ ^1 \Sigma_u^+$ & -0.629$^c$ & -0.628 & -0.625 \\
      \bottomrule
    \end{tabular}

  \end{center}

  \footnotesize
  $^a$\citeasnoun{Kolos1986}; $^b$\citeasnoun{Staszewska1999}; $^c$\citeasnoun{Staszewska2002};
  $^d$\citeasnoun{Wolniewicz1994}; $^e$\citeasnoun{Wolniewicz2003}; $^f$This work;
  $^g$\citeasnoun{Zammit2017a}.
\end{table}

\subsubsection{Scattering Model}%
\label{ssub:scattering_model}

In the R-matrix method the electronic density of the target must be contained within the R-matrix
sphere, which is of radius $a$. Due to the extremely diffuse nature of our basis set we used a
radius of $a=100 \ a_0$. Usually an R-matrix sphere of this size would be impossible, as the
continuum basis set required to fill the space would suffer from severe linear dependence. However,
as mentioned previously, the new UKRMol+ codes allow the use of BTOs which are numerically stable
regardless of the size of the R-matrix sphere. We found that, for molecular hydrogen, using a BTO
only continuum basis not only removed linear dependence issues but it also gave a better description
of the continuum. Details of the continuum basis can be found in Table \ref{tab:continuum_basis}.

\begin{table}[t]
  \caption{ \textbf{Continuum Basis} Parameters for the continuum basis.}
  \vspace{5pt}
  \centering
  \begin{tabular}{ll}
    \toprule
    Property                      & Value \\
    \midrule
    Number of B-Splines (per $l$) & 75    \\
    B-spline Order                & 9     \\
    $l_{\mathrm{max}}$            & 6     \\
    \bottomrule
  \end{tabular}
  \label{tab:continuum_basis}
\end{table}

To solve the scattering problem we are using a close-coupling expansion. This is necessary for
describing exchange and polarisation effects in addition to modelling electronic excitation. To
ensure balance between the $N+1$ and $N$-electron contributions in the close-coupling expansion we
use a similar treatment in the $N+1$ electron system as we did for the target. We adopt two types of
configuration state function (CSF) in the $N+1$ system. There are those where one electron occupies
a continuum orbital, and those where all of the $N+1$ electrons occupy the target molecular
orbitals. This amounts to;
\begin{eqnarray}
& (target)^2 (continuum)^1, \nonumber \\
& (target)^3, \nonumber
\end{eqnarray}
where $target$ stands for the complete set of target molecular orbitals. Note that in the first
configuration step it is necessary to couple the target electrons to the appropriate symmetry, in
order to facilitate the identification of the correct target states \cite{jt180}. However, there are
no constraints on the ``$L^2$'' configurations generated in the second step. Using this model we
retain all the target states below 30 eV vertical excitation energy, which is 98 states. This
generates an average of 65,000 CSFs per molecular symmetry.

Up to now differential cross-sections (DCS) obtained from R-matrix calculations were generated using
the program POLYDCS \cite{Sanna1998} which includes rotational excitation of the molecule but is
limited to electronically elastic transitions. Therefore we have developed a new program for the
calculation of DCS which includes only orientational averaging of the molecule but is applicable to
electronically inelastic transitions and optionally employs the standard top-up procedure based on
the first Born approximation for inelastic dipolar scattering. For details
see~\ref{sec:top_up_methods}.

\subsection{Molecular convergent close-coupling}%
\label{sub:convergent_close_coupling}

The MCCC method is a momentum-space formulation of the close-coupling theory. The target spectrum is
represented by a set of (pseudo)states generated by diagonalising the target electronic Hamiltonian
in a basis of Sturmian (Laguerre) functions. For a suitable choice of basis the resulting states
provide a sufficiently accurate representation of the low-lying discrete spectrum and a
discretisation of the continuous spectrum, which allows the effects of coupling to ionisation
channels to be modelled. Expanding the total scattering wave function in terms of the target
pseudostates and performing a partial-wave expansion of the projectile wave function leads to a set
of linear integral equations for the partial-wave $T$-matrix elements, which are solved using
standard techniques. The strength of the MCCC method is the ability to perform calculations with
very large close-coupling expansions, allowing for the explicit demonstration of convergence in the
scattering quantities of interest with respect to the number of target states included in the
calculations and the size of the projectile partial-wave expansion.

The MCCC method has been implemented for electron and positron scattering on diatomic molecules in
both spherical and spheroidal coordinates. The spherical implementation is simpler and provides an
adequate description of the molecular structure at the mean internuclear separation of the H$_2$
ground state. We have utilised the spherical MCCC method for detailed convergence studies and the
calculation of elastic, excitation, ionisation, and grand-total cross-sections over a wide range of
incident energies. Spheroidal coordinates are a more natural system for describing the electronic
structure at larger $R$, where the target wave functions become more diffuse. We have utilised the
spheroidal MCCC method to calculate vibrationally-resolved cross-sections for excitation of a number
of low-lying states of H$_2$, including scattering on all bound vibrational levels of the ground
electronic state \cite{Scarlett19adndt}. This has allowed detailed studies to be performed for
dissociation of H$_2$ in the ground and vibrationally-excited states
\cite{atoms7030075,TSSFZB18,Tapley18,Scarlett_diss2018}, and vibrational excitation of the
$X~^1\Sigma_g^+$ state via electronic excitation and radiative decay \cite{Scarlett_2019}. For
clarity of presentation, in the present paper we present only the spherical MCCC results. For
details of the spheroidal MCCC method and comparisons of the spherical and spheroidal MCCC cross
sections see \citeasnoun{posmolCCC19}.

\subsubsection{Target Model}%
\label{ssub:CCC_target_model}

The MCCC target structure is obtained using a CI calculation. The basis for the CI expansion
consists of two-electron configurations formed by products $(n\ell,n'\ell')$ of one-electron
Laguerre-based orbitals. To reduce the number of two-electron states generated, we allow one of the
target electrons to occupy any one-electron orbital, while the other is restricted to the $1s$,
$2s$, and $2p$ orbitals. The largest target structure calculation we have performed utilises a
Laguerre basis of $N_\ell=17-\ell$ functions with $\ell\leq3$, which generates a total of 491
states. To improve the accuracy of the $X~^1\Sigma_g^+$ and $b~^3\Sigma_u^+$ states, where the
multicentre effects are strongest, we replace the $1s$ Laguerre function with an accurate H$_2^+$
$1s\sigma_g$ state obtained via diagonalisation of the H$_2^+$ Hamiltonian in a basis with
$N_\ell=60-\ell$ functions for $\ell\leq8$.

\subsubsection{Scattering Models}%
\label{ssub:CCC_scattering_models}

Fixed-Nuclei (FN) MCCC calculations were performed at $R=1.448 \ a_0$ using a number of scattering
models, ranging from 9 to 491 states included in the close-coupling expansion. This allowed for a
detailed investigation of convergence and the effects of including various reaction channels (see
\citeasnoun{Zammit2017a} for details). The MCCC results presented here were obtained from the
491-state model, which yielded convergent cross-sections for each of the transitions of interest.
With regards to the partial-wave expansion of the projectile wave function, we have included angular
momenta up to $L_\mathrm{max}=8$, and all total angular momentum projections up to
${M_\mathrm{max}=L_\mathrm{max}}$. To account for the contributions from higher partial waves we
utilise an analytical Born subtraction (ABS) technique, which is equivalent to replacing the
$L>L_\mathrm{max}$ cross-sections with their respective partial-wave Born cross-sections. We have
found that the partial-wave expansion with $L_\mathrm{max}=8$ produces convergent integrated cross
sections (ICS) for all transitions considered here when used in conjunction with the ABS technique.
For dipole-allowed transitions, the partial-wave convergence of the DCS can be considerably slower
than it is for the ICS. The method we have adopted to resolve this issue is discussed in
\citeasnoun{Zammit2017a}. For the $X~^1\Sigma_g^+\rightarrow b~^3\Sigma_u^+$ transition,
adiabatic-nuclei calculations have been performed at low incident energies using a model consisting
of 12 target states which yields convergent cross-sections for the $b~^3\Sigma_u^+$ state below
approximately 15~eV. These calculations are described in \citeasnoun{Scarlett17b3}.

\subsection{Adiabatic-Nuclei Approximation}%
\label{sub:adiabatic_nuclei_approximation}

So far we have discussed FN calculations. However, in reality, the molecular geometry is not fixed
and experiment effectively samples from a range of initial and final states. This will have an
impact on both the integrated and differential cross-sections. This behaviour is most notable near
threshold \cite{jt229}. At higher scattering energies, away from the threshold, the two
approximations converge as nuclear motion effects become less significant. We use the
Adiabatic-Nuclei (AN) approach detailed in \citeasnoun{Lane1980} which has been recently
demonstrated by \citeasnoun{Scarlett17b3} on molecular hydrogen. In this work we use the ground
vibrational wavefunction to vibrationally average multiple FN calculations, carried out at a range
of different nuclear geometries. Although, in general, this method can also be used to produce
vibrationally resolved cross-sections.

\begin{table}[ht]
  \caption{ Absolute target energies, E (a.u.), and vertical excitation energies, $\Delta$E (eV), at
  the mean vibrational bond length $R_0=1.448 \ a_0$. RM data are from UKRMol+ (this work) and MCCC
  from \citeasnoun{Zammit2017a}.}
  \vspace{5pt}
  \centering
  \begin{tabular}{rccccc}
    \toprule
                          & \multicolumn{2}{c}{E (a.u.)} & & \multicolumn{2}{c}{$\Delta$E (eV)} \\
                          \cline{2-3} \cline{5-6}
    State                 & RM     & MCCC    & & RM    & MCCC \\
    \midrule
    $X   \ ^1 \Sigma_g^+$ & -1.172 & -1.161 & &     - &     - \\
    $b   \ ^3 \Sigma_u^+$ & -0.796 & -0.782 & & 10.23 & 10.31 \\
    $a   \ ^3 \Sigma_g^+$ & -0.718 & -0.715 & & 12.35 & 12.14 \\
    $B   \ ^1 \Sigma_u^+$ & -0.712 & -0.704 & & 12.52 & 12.44 \\
    $c   \ ^3 \Pi_u$      & -0.712 & -0.707 & & 12.52 & 12.35 \\
    $EF  \ ^1 \Sigma_g^+$ & -0.697 & -0.693 & & 12.93 & 12.73 \\
    $C   \ ^1 \Pi_u$      & -0.694 & -0.693 & & 13.02 & 12.73 \\
    $e   \ ^3 \Sigma_u^+$ & -0.650 & -0.647 & & 14.21 & 13.99 \\
    $h   \ ^3 \Sigma_g^+$ & -0.636 & -0.634 & & 14.60 & 14.34 \\
    $B'  \ ^1 \Sigma_u^+$ & -0.635 & -0.631 & & 14.63 & 14.42 \\
    $d   \ ^3 \Pi_u$      & -0.634 & -0.632 & & 14.65 & 14.39 \\
    \bottomrule
  \end{tabular}

  \label{tab:target_energies_R0}
\end{table}

\section{Results}%
\label{sec:results}

In this section, we present FN ICS and DCS for elastic and inelastic processes. For inelastic
processes we consider the first ten electronic excited states. In the second section we use the
adiabatic-nuclei approximation to introduce nuclear motion effects which are particularly important
close to threshold. The FN R-matrix ICS and DCS data are provided as supplementary data.

The scattering calculations that follow were carried out at the mean vibrational bond length,
$R_0=1.448 \ a_0$, to provide the best comparison to experiment, within the FN approximation. Table
\ref{tab:target_energies_R0} lists the target states and the vertical excitation energies obtained
for both methods. Similarly, compared to Table \ref{tab:target_energies_Re}, the R-matrix target
energies are more accurate than the MCCC method, as they are lower in energy (note that both methods
are variational). However, it should be noted that the absolute energy is of less significance for
this work, and that the vertical excitation energies (relative to the ground state) are in good
agreement.

\subsection{Fixed-Nuclei Cross-Sections}
\label{sub:fixed_nuclei}

We present ICS and DCS for the first ten target states (see Table \ref{tab:target_energies_R0}); $X
\ ^1\Sigma_g^+$, $b \ ^3\Sigma_u^+$, $a \ ^3\Sigma_g^+$, $B \ ^1\Sigma_u^+$, $c \ ^3\Pi_u$, $EF \
^1\Sigma_g^+$, $C \ ^1\Pi_u$, $e \ ^3\Sigma_u^+$, $h \ ^3\Sigma_g^+$, $B' \ ^1\Sigma_u^+$ and $d \
^3\Pi_u$. Where available, recommended cross-sections and experimental results are plotted against
the two theoretical calculations.

\begin{figure}[htbp]
  \centering
  \includegraphics[width=1.0\linewidth]{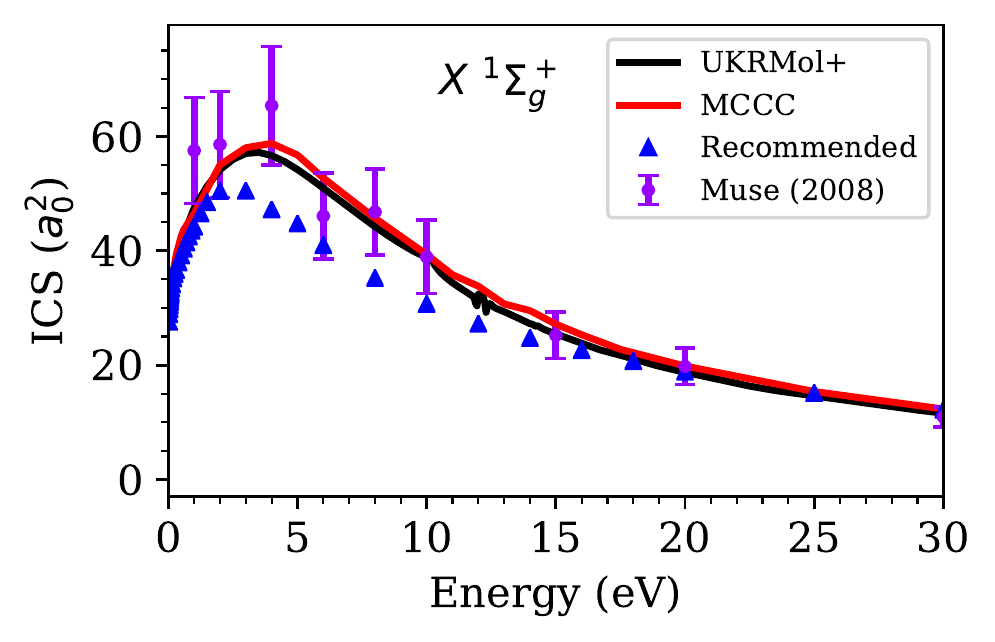}
  \caption{ ICS for elastic collisions. Comparison of the UKRMol+ and MCCC calculations with the
  measurements of \citeasnoun{Muse2008} and recommended data of \citeasnoun{Yoon2008}.}
  \label{fig:ics_state0}
\end{figure}

\begin{figure}[htbp]
  \centering
  \includegraphics[width=1.0\linewidth]{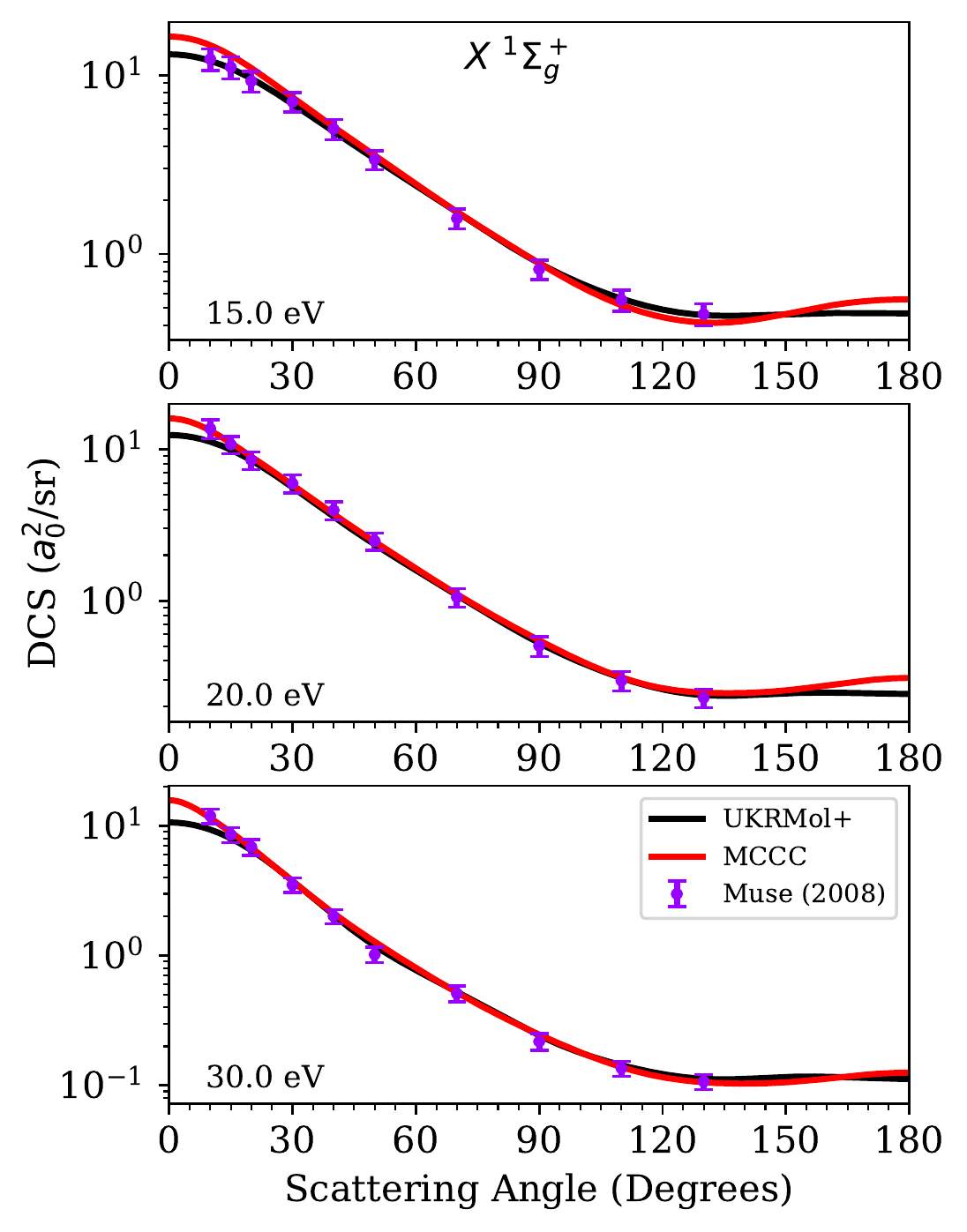}
  \caption{ DCS for elastic collisions. Comparison of the UKRMol+ and MCCC calculations with the
  measurements of \citeasnoun{Muse2008}.}
  \label{fig:dcs_state0}
\end{figure}

\subsubsection{Elastic Cross-Sections}
\label{ssub:elastic}

The elastic ICS (Fig.~\ref{fig:ics_state0}) demonstrates good agreement between MCCC and R-matrix
theory. The calculated data lie within the error bars of the experiment conducted by
\citeasnoun{Muse2008}. For the DCS (Fig. \ref{fig:dcs_state0}) at scattering angles exceeding
15\dgr\ the two theories essentially overlap. At energies greater than 15 eV the R-matrix
calculations have a diminished forward peak and this is due to a lack of convergence of the partial
wave expansion. Due to computational constraints $L_\mathrm{max}=6$ for the R-matrix calculations.
This compares to $L_\mathrm{max}=8$ for the MCCC calculation, which also employs the ABS technique.
Nevertheless, scattering angles close to $\theta=$ 0 or $\theta=$ 180 do not contribute as much to
the ICS due to a $\sin \theta$ term in the integrand. Therefore, despite the differences in the DCSs
the resulting ICSs are similar.

The recommended data of \citeasnoun{Yoon2008} for the ICS are noticeably lower than those obtained
from the R-matrix and MCCC calculations (Fig. \ref{fig:ics_state0}). Whilst they are within their
specified margin of error ($\pm 20\%$) we believe that, due to the excellent agreement between both
theories and experiment for the DCS (Fig. \ref{fig:dcs_state0}), the recommended data should be
revised.

\subsubsection{Triplet States}
\label{ssub:triplet_states}

\begin{figure}[htbp]
  \centering
  \includegraphics[width=1.0\linewidth]{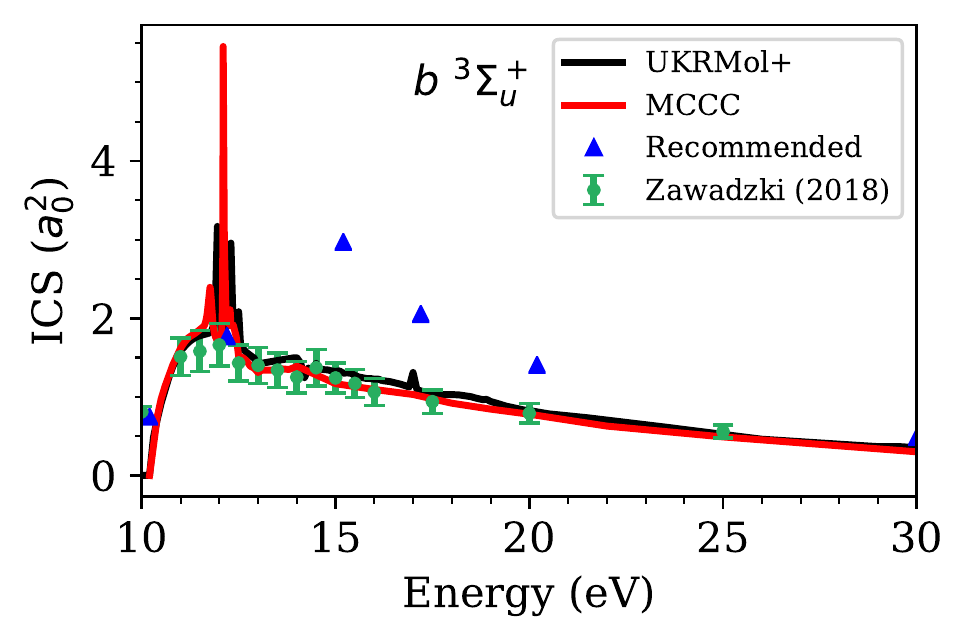}
  \caption{ ICS for the $X \ ^1\Sigma_g^+ \rightarrow b \ ^3\Sigma_u^+ $ transition. Comparison of
  the UKRMol+ and MCCC calculations with the measurements of \citeasnoun{Zawadzki2018} and
  recommended data of \citeasnoun{Yoon2008}.}
  \label{fig:ics_state1}
\end{figure}

The first excited electronic state is $b \ ^3\Sigma_u^+$. For this state we have used a fine energy
grid for both the MCCC and UKRMol+ calculations. This allows an accurate comparison of the two ICSs.
In Fig. \ref{fig:ics_state1} prominent resonance structures are observed near 12 eV. Across the
energy range considered the two calculations agree.

\begin{figure}[htbp]
  \centering
  \includegraphics[width=1.0\linewidth]{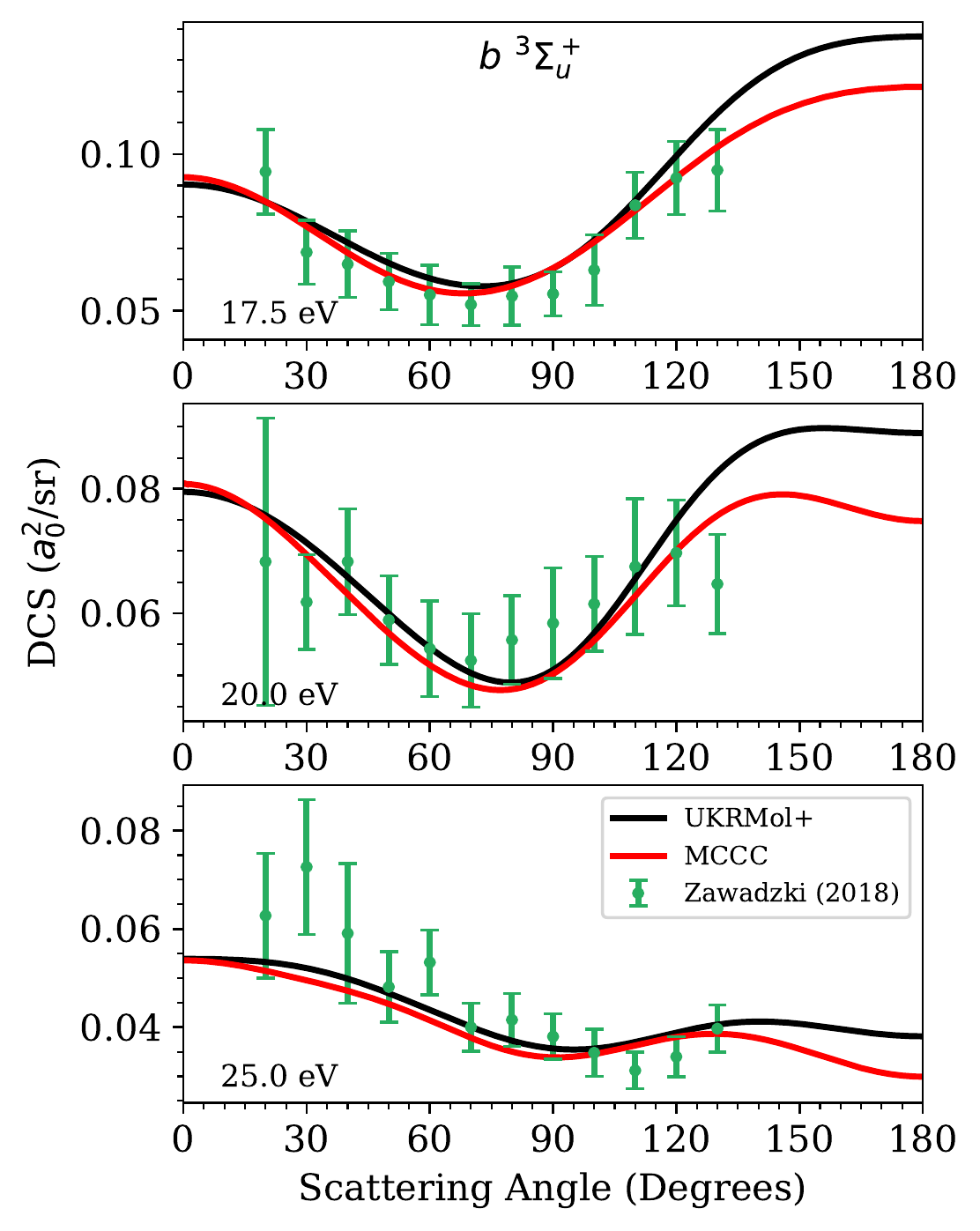}
  \caption{ DCS for the $X \ ^1\Sigma_g^+ \rightarrow b \ ^3\Sigma_u^+ $ transition. Comparison of
  the UKRMol+ and MCCC calculations with the measurements of \citeasnoun{Zawadzki2018}.}
  \label{fig:dcs_state1}
\end{figure}

The current recommended cross-sections agree at low energy but from 15 eV to 20 eV they appear to
overestimate the cross-section. The newer experiment from \citeasnoun{Zawadzki2018} is much closer
to the two theories. The DCSs (Fig. \ref{fig:dcs_state1}) also agree closely with these experimental
data. The R-matrix calculations are a little higher than the MCCC calculations for angles exceeding
135$^\circ$ but, again, the affect on the ICS is insignificant.

\begin{figure}[htbp]
  \centering
  \includegraphics[width=1.0\linewidth]{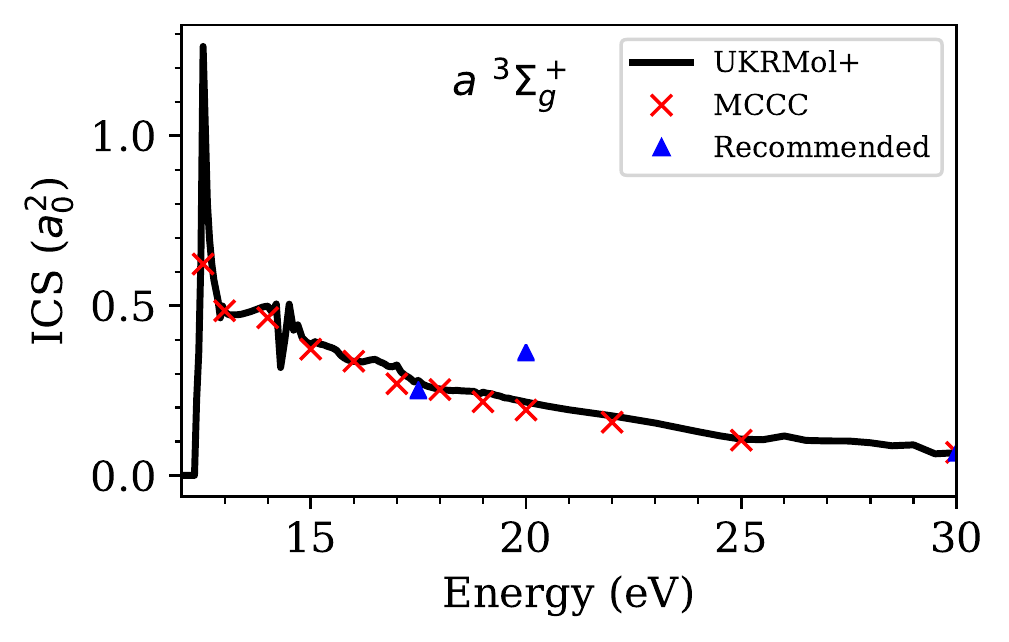}
  \caption{ ICS for the $X \ ^1\Sigma_g^+ \rightarrow a \ ^3\Sigma_g^+ $ transition. Comparison of
  the UKRMol+ and MCCC calculations with the recommended data of \citeasnoun{Yoon2008}.}
  \label{fig:ics_state2}
\end{figure}

\begin{figure}[htbp]
  \centering
  \includegraphics[width=1.0\linewidth]{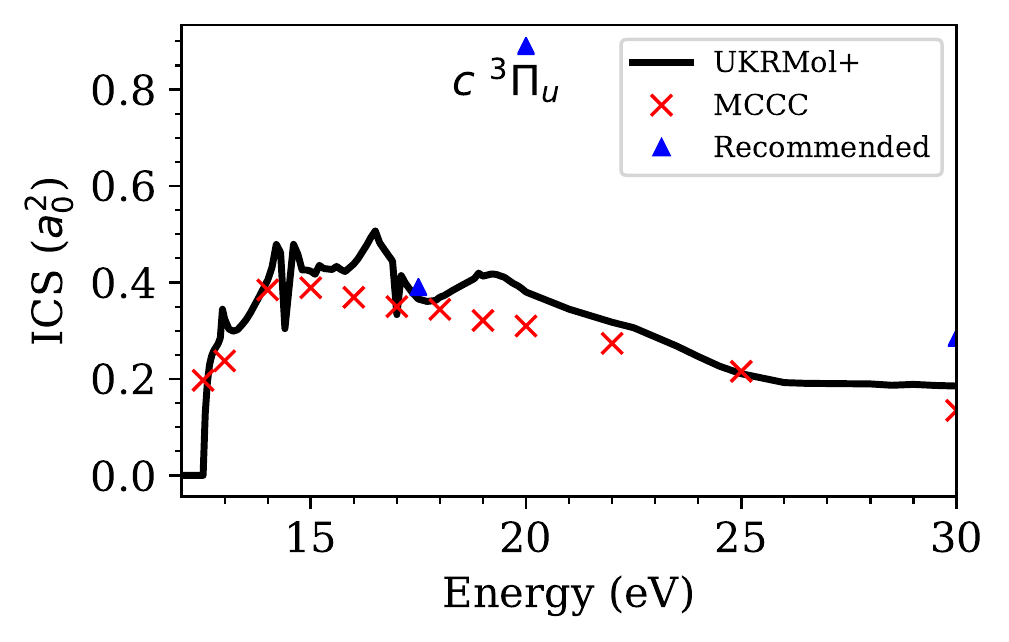}
  \caption{ ICS for the $X \ ^1\Sigma_g^+ \rightarrow c \ ^3\Pi_u $ transition. Comparison of the
  UKRMol+ and MCCC calculations with the recommended data of \citeasnoun{Yoon2008}.}
  \label{fig:ics_state4}
\end{figure}

\begin{figure}[htbp]
  \centering
  \includegraphics[width=1.0\linewidth]{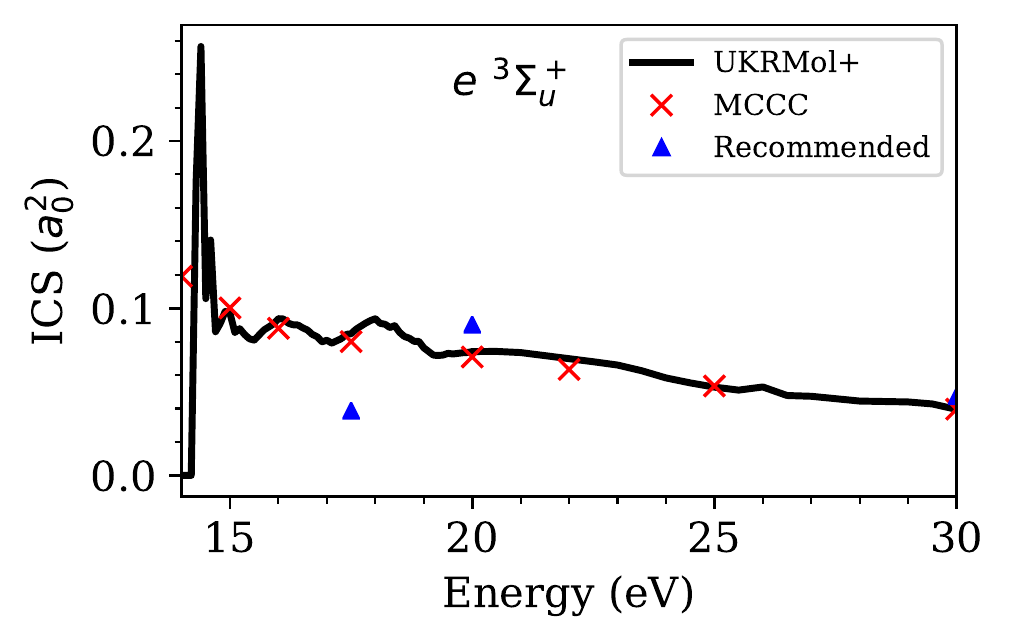}
  \caption{ ICS for the $X \ ^1\Sigma_g^+ \rightarrow e \ ^3\Sigma_u^+ $ transition. Comparison of
  the UKRMol+ and MCCC calculations with the recommended data of \citeasnoun{Yoon2008}.}
  \label{fig:ics_state7}
\end{figure}

For higher excited states i.e., those energetically above $b \ ^3\Sigma_u^+$, MCCC results are
presented on a coarser energy grid. Therefore, we can no longer compare the narrow resonant
structures. The ICSs for states $a \ ^3\Sigma_g^+$, $c \ ^3\Pi_u$ and $e \ ^3\Sigma_u^+$ (Figs.
\ref{fig:ics_state2}, \ref{fig:ics_state4} and \ref{fig:ics_state7} respectively) have good
agreement between the MCCC and R-matrix theories.

The recommended data points are based on the EELS (Electron Energy Loss Spectroscopy) experiment of
\citeasnoun{Wrkich2002}. The data points are sparse so it is hard to quantitatively compare against
the two theory calculations. However, given agreement between the two theoretical calculations and
more recent experiments, we believe that the recommended cross-sections should be revised for all of
the triplet states considered so far.

\begin{figure}[htbp]
  \centering
  \includegraphics[width=1.0\linewidth]{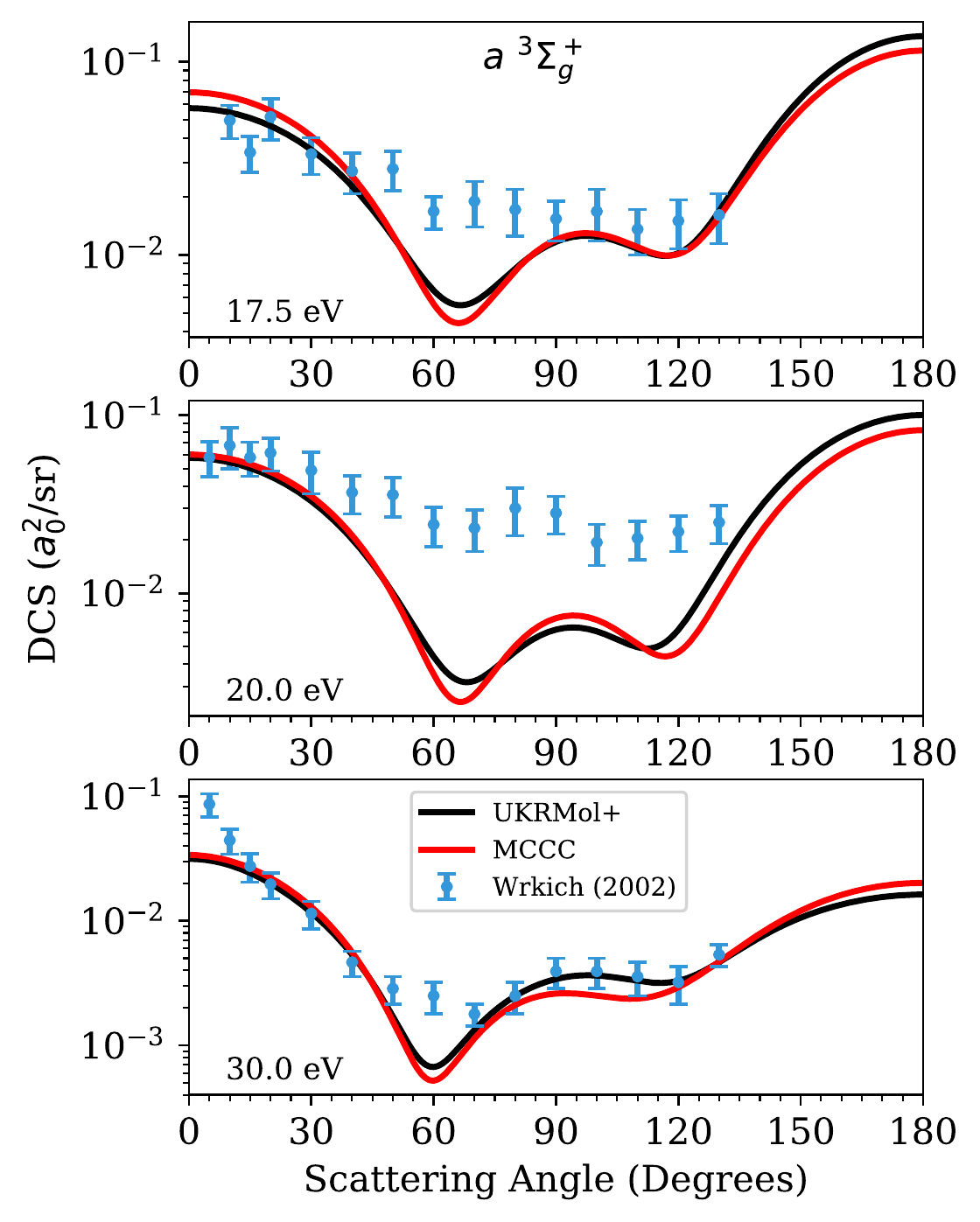}
  \caption{ DCS for the $X \ ^1\Sigma_g^+ \rightarrow a \ ^3\Sigma_g^+ $ transition. Comparison of
  the UKRMol+ and MCCC calculations with the measurements of \citeasnoun{Wrkich2002}.}
  \label{fig:dcs_state2}
\end{figure}

The DCSs shed more light on the comparison. Fig. \ref{fig:dcs_state2} shows the $a \ ^3\Sigma_g^+$
state. Agreement is best for 17.5 eV and 30 eV. The general shape is present at all three energies.
That is, the cross-section dips around 60\dgr\ and 120\dgr. However, for intermediate angles the
magnitude of the DCS is higher (especially for 20 eV) than the theoretical calculations. EELS
experiments are hard to conduct for excited states of H$_2$ because the states overlap in the
spectra and the individual components have to be deconvoluted. Based on the difficulty of these type
of experiments for highly-excited states we suggest that the calculations are more reliable.

\begin{figure}[htbp]
  \centering
  \includegraphics[width=1.0\linewidth]{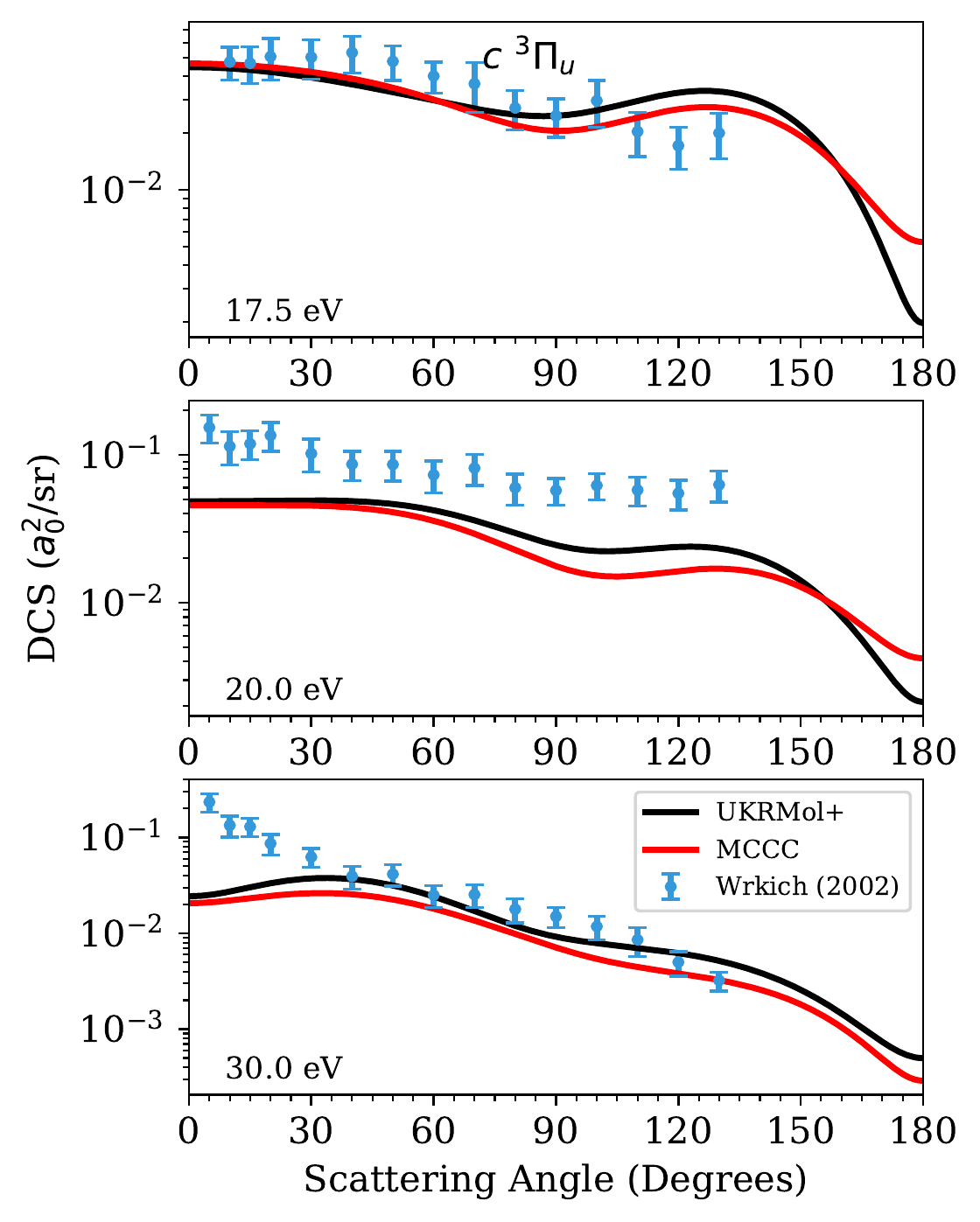}
  \caption{ DCS for the $X \ ^1\Sigma_g^+ \rightarrow c \ ^3\Pi_u $ transition. Comparison of the
  UKRMol+ and MCCC calculations with the measurements of \citeasnoun{Wrkich2002}.}
  \label{fig:dcs_state4}
\end{figure}

For the $c \ ^3\Pi_u$ state (Fig. \ref{fig:dcs_state4}) the situation is similar to the $a \
^3\Sigma_g^+$ state. There is a slight downward slope towards higher scattering angles that is
present in both the calculation and the experiment. However, the experimental DCS at 20 eV is
approximately an order of magnitude higher.

\begin{figure}[htbp]
  \centering
  \includegraphics[width=1.0\linewidth]{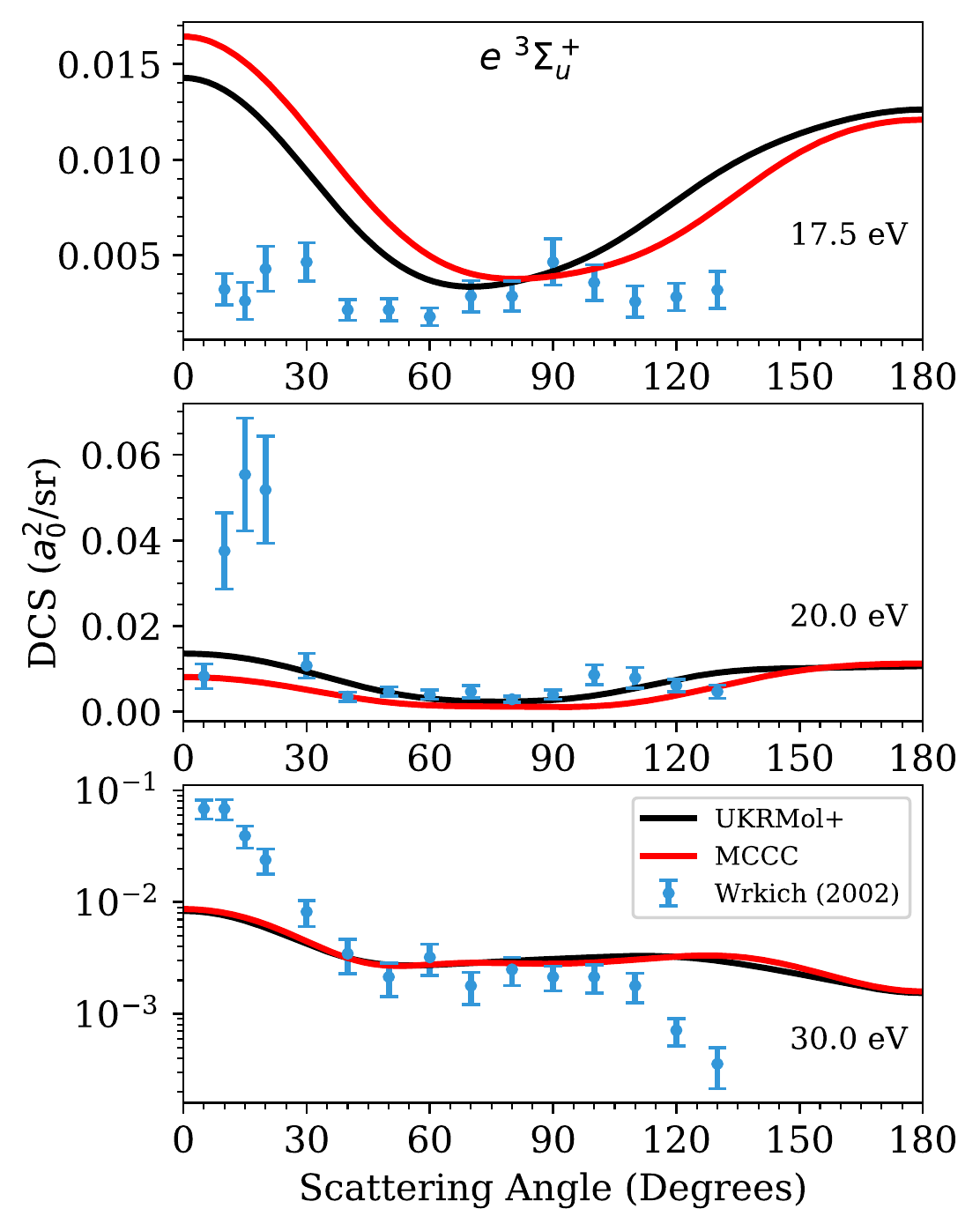}
  \caption{ DCS for the $X \ ^1\Sigma_g^+ \rightarrow e \ ^3\Sigma_u^+ $ transition. Comparison of
  the UKRMol+ and MCCC calculations with the measurements of \citeasnoun{Wrkich2002}.}
  \label{fig:dcs_state7}
\end{figure}

For the $e \ ^3\Sigma_u^+$ state there is no qualitative agreement between theory and experiment. At
all three energies (shown in Fig. \ref{fig:dcs_state7}) we have large discrepancies for low angle
scattering i.e., below 30\dgr. This is not too surprising though as low and high angle scattering is
difficult to measure due to the physical constraints of the experimental setup. Therefore, we
suspect that the low angle cross-sections measured by \citeasnoun{Wrkich2002}, at 20 eV and 30 eV,
are too high.

\begin{figure}[htbp]
  \centering
  \includegraphics[width=1.0\linewidth]{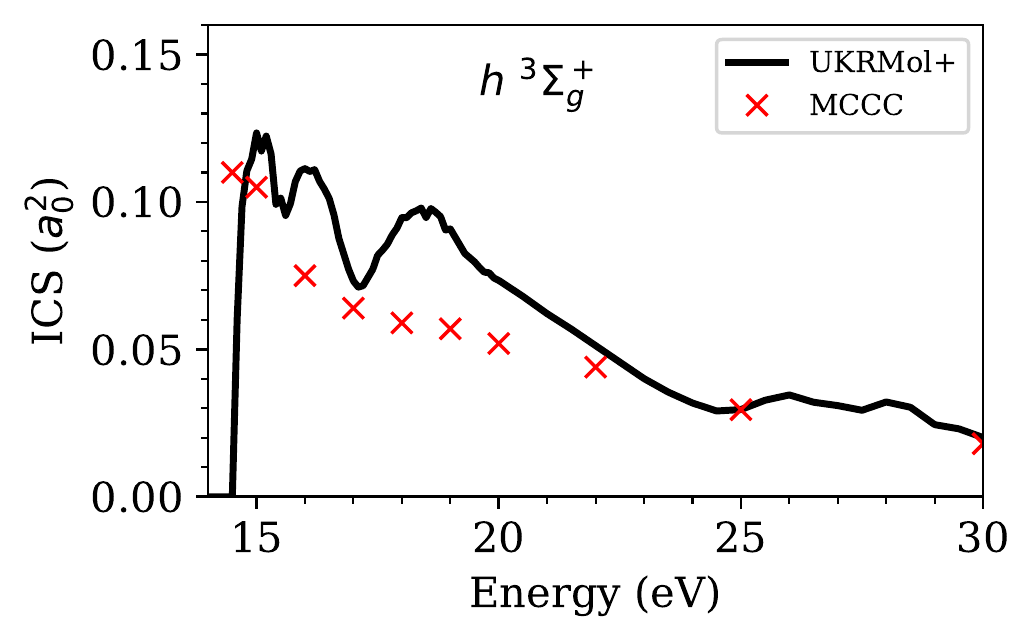}
  \caption{ ICS for the $X \ ^1\Sigma_g^+ \rightarrow h \ ^3\Sigma_g^+ $ transition. Comparison of
  the UKRMol+ and MCCC calculations.}
  \label{fig:ics_state8}
\end{figure}

\begin{figure}[htbp]
  \centering
  \includegraphics[width=1.0\linewidth]{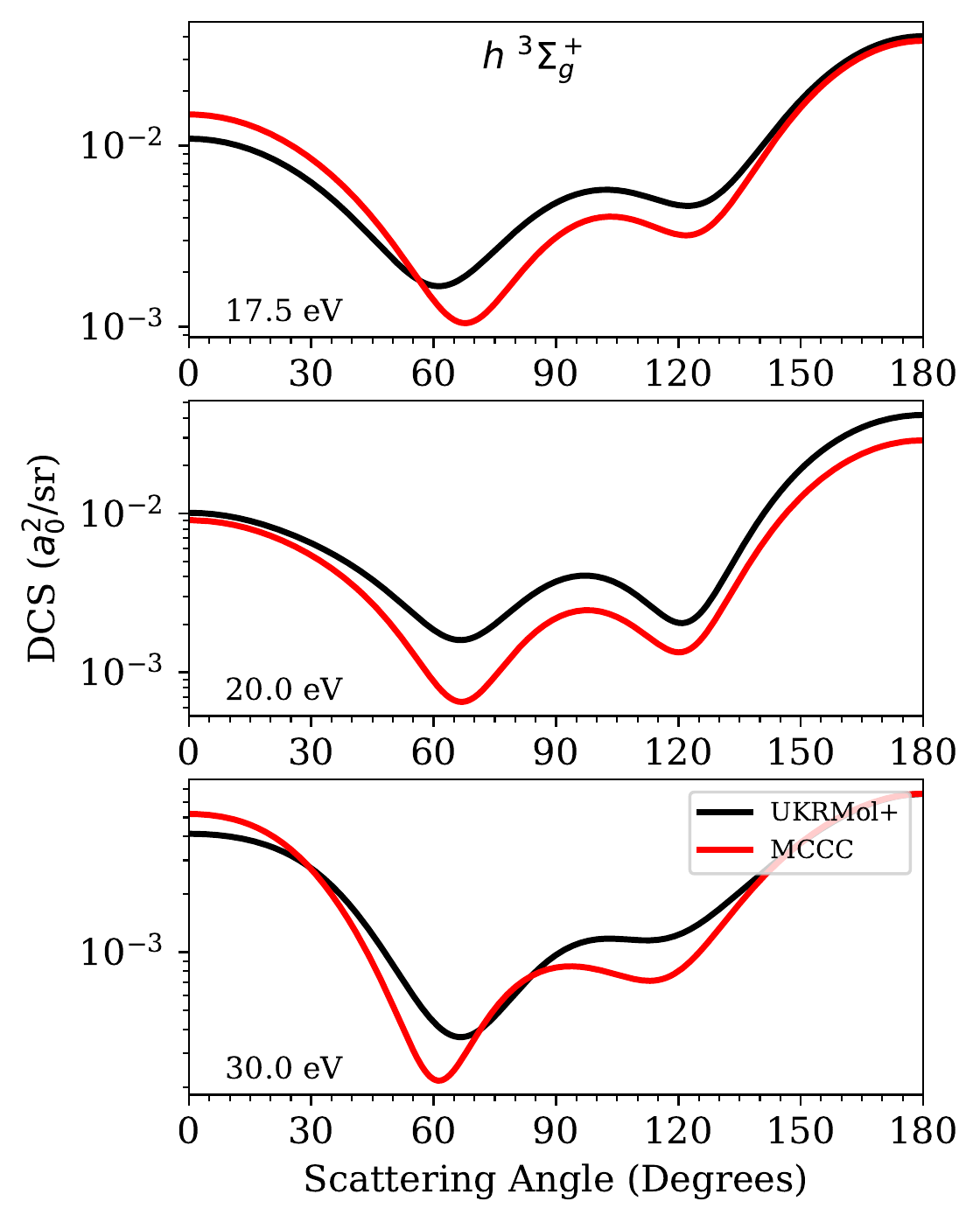}
  \caption{ DCS for the $X \ ^1\Sigma_g^+ \rightarrow h \ ^3\Sigma_g^+ $ transition. Comparison of
  the UKRMol+ and MCCC calculations.}
  \label{fig:dcs_state8}
\end{figure}

\begin{figure}[htbp]
  \centering
  \includegraphics[width=1.0\linewidth]{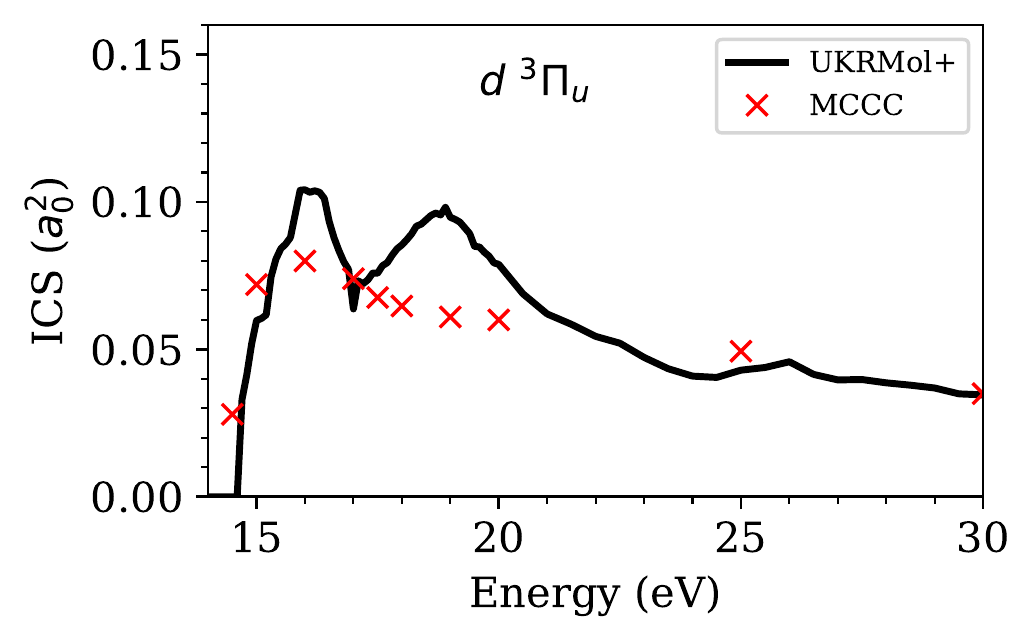}
  \caption{ ICS for the $X \ ^1\Sigma_g^+ \rightarrow d \ ^3\Pi_u $ transition. Comparison of the
  UKRMol+ and MCCC calculations.}
  \label{fig:ics_state10}
\end{figure}

The ICSs for states $h \ ^3\Sigma_g^+$ and $d \ ^3\Pi_u$ (Figs. \ref{fig:ics_state8} and
\ref{fig:ics_state10}) show agreement between the two theories. However, the R-matrix calculation
exhibits pronounced features around 16 eV and 19 eV. In the standard R-matrix approach used in this
work, ionisation effects are not included. We include states above the ionisation threshold but we
do not explicitly include pseudostates. To model ionisation pseudostates are required as implemented
in the R-matrix with pseudostates (RMPS) method \cite{Gorfinkiel2005}. As a result, the
cross-section is overestimated above ionisation threshold. This behaviour was demonstrated
previously in MCCC calculations when only the bound states were used \cite{Zammit2017a}. In
addition, weak transitions can also suffer from small oscillations, but the impact is reduced as the
size of the close-coupling expansion increases. Therefore, the enhanced R-matrix cross-section is
likely due to missing ionisation channels.

\begin{figure}[htbp]
  \centering
  \includegraphics[width=1.0\linewidth]{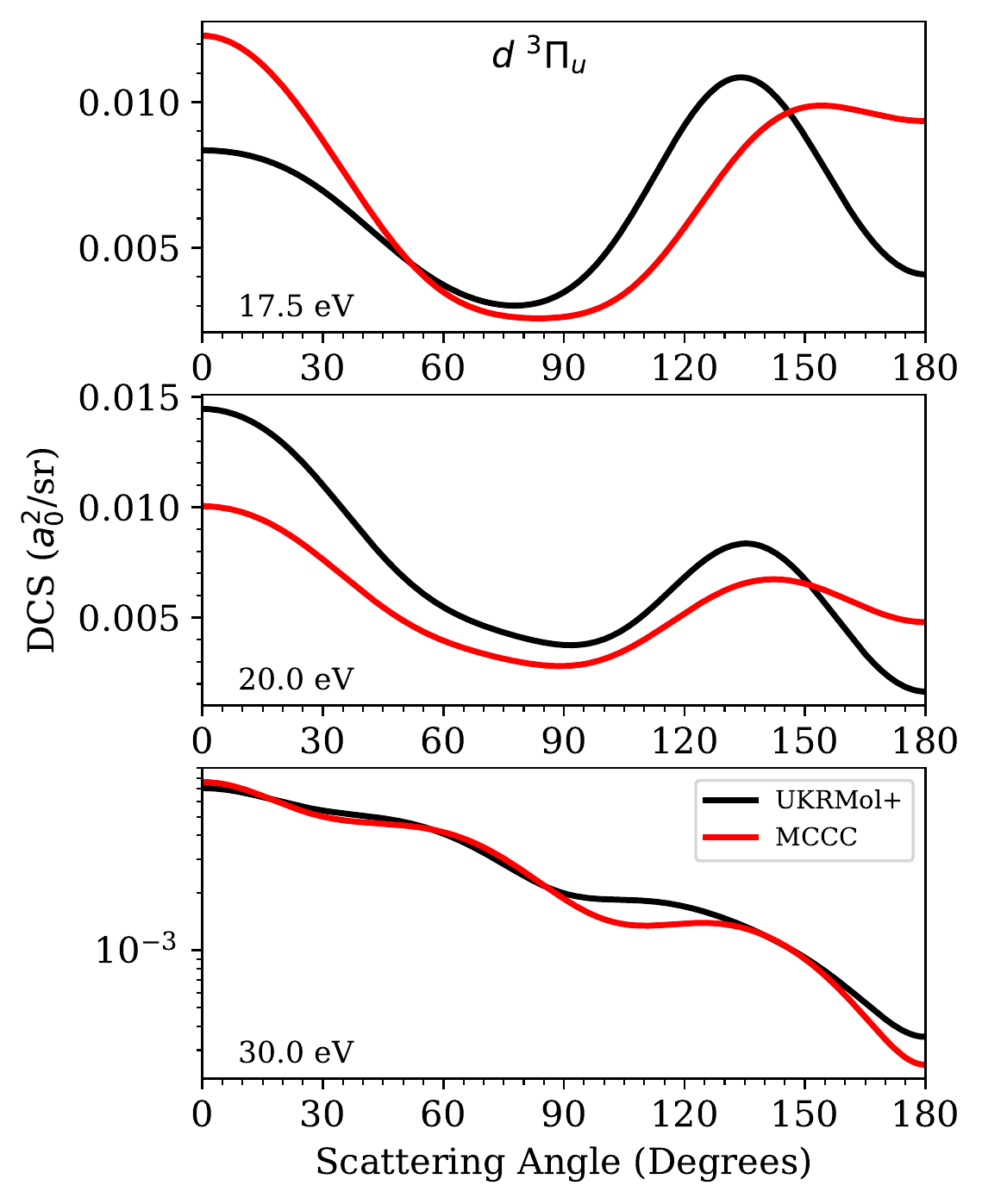}
  \caption{ DCS for the $X \ ^1\Sigma_g^+ \rightarrow d \ ^3\Pi_u $ transition. Comparison of the
  UKRMol+ and MCCC calculations.}
  \label{fig:dcs_state10}
\end{figure}

The R-matrix DCSs for state $h \ ^3\Sigma_g^+$ (Fig. \ref{fig:dcs_state8}) show broad agreement with
the MCCC data. There are no recommended data for either the $h \ ^3\Sigma_g^+$ or $d \ ^3\Pi_u$
states. For the $d \ ^3\Pi_u$ state (Fig. \ref{fig:dcs_state10}) there are more significant
differences between the two theories. As the target excitation increases, we typically expect less
agreement between the two theories. Higher excited states tend to be less accurately described by
the electronic structure calculations used in the R-matrix method.

\clearpage
\subsubsection{Singlet States}
\label{ssub:all}

\begin{figure}[htbp]
  \centering
  \includegraphics[width=1.0\linewidth]{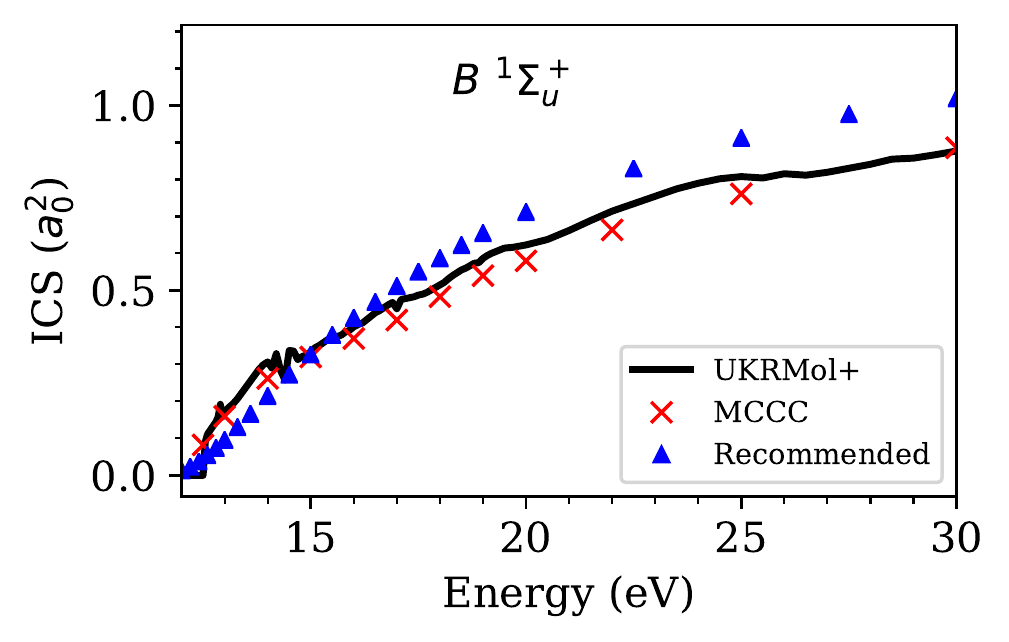}
  \caption{ ICS for the $X \ ^1\Sigma_g^+ \rightarrow B \ ^1\Sigma_u^+ $ transition. Comparison of
  the UKRMol+ and MCCC calculations with the recommended data of \citeasnoun{Yoon2008}.}
  \label{fig:ics_state3}
\end{figure}

\begin{figure}[htbp]
  \centering
  \includegraphics[width=1.0\linewidth]{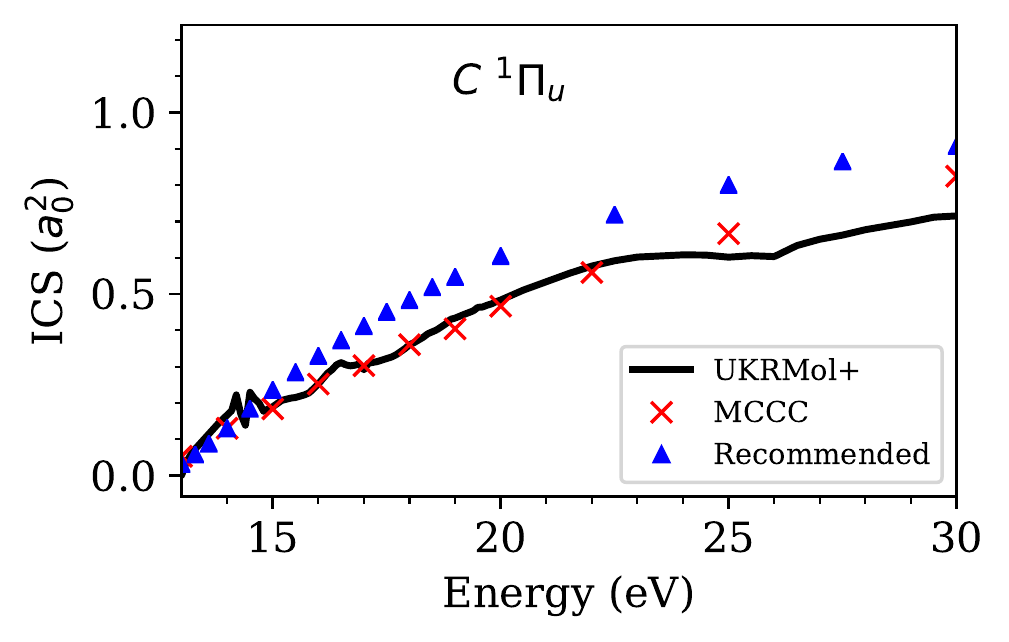}
  \caption{ ICS for the $X \ ^1\Sigma_g^+ \rightarrow C \ ^1\Pi_u $ transition. Comparison of the
  UKRMol+ and MCCC calculations with the recommended data of \citeasnoun{Yoon2008}.}
  \label{fig:ics_state6}
\end{figure}

\begin{figure}[htbp]
  \centering
  \includegraphics[width=1.0\linewidth]{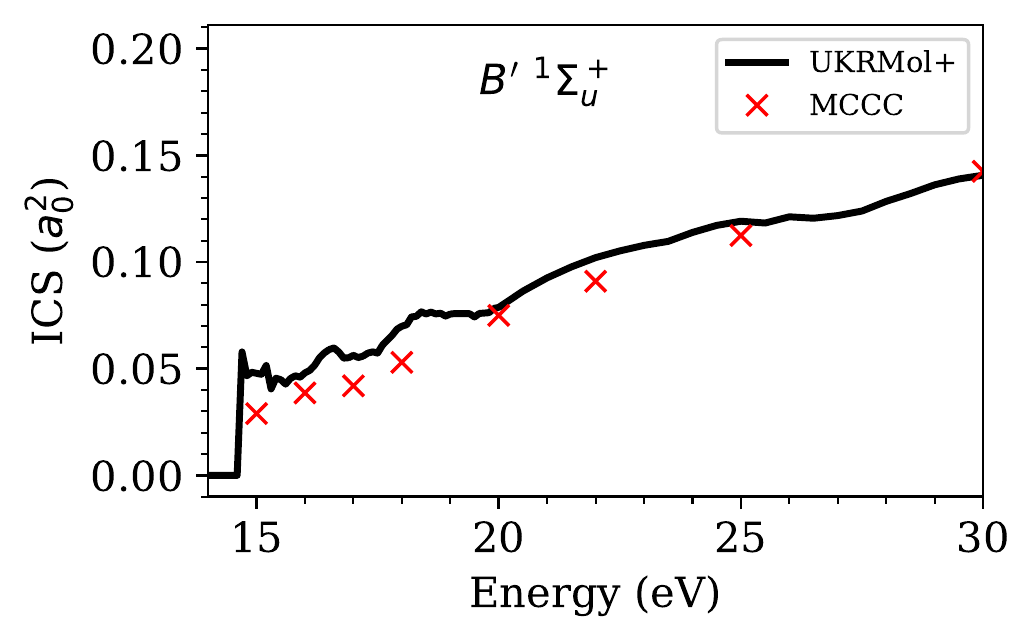}
  \caption{ ICS for the $X \ ^1\Sigma_g^+ \rightarrow B^\prime \ ^1\Sigma_u^+ $ transition.
  Comparison of the UKRMol+ and MCCC calculations.}
  \label{fig:ics_state9}
\end{figure}

Next we consider the singlet states. ICSs for three dipole-allowed states, $B \ ^1\Sigma_u^+$, $C \
^1\Pi_u$ and $B^\prime \ ^1\Sigma_u^+$, are shown in Figs. \ref{fig:ics_state3},
\ref{fig:ics_state6} and \ref{fig:ics_state9} respectively. All three ICSs show excellent agreement
between MCCC and R-matrix theory. Furthermore, agreement with the recommended data, which is
available for the $B \ ^1\Sigma_u^+$ and $C \ ^1\Pi_u$ states, is extremely good at the energies
considered here. Contrary to the previous EELS experimental data, these recommended data were
measured from the optical emission of the electron impact electronically excited B and C states
\cite{Liu1998}.

\begin{figure}[htbp]
  \centering
  \includegraphics[width=1.0\linewidth]{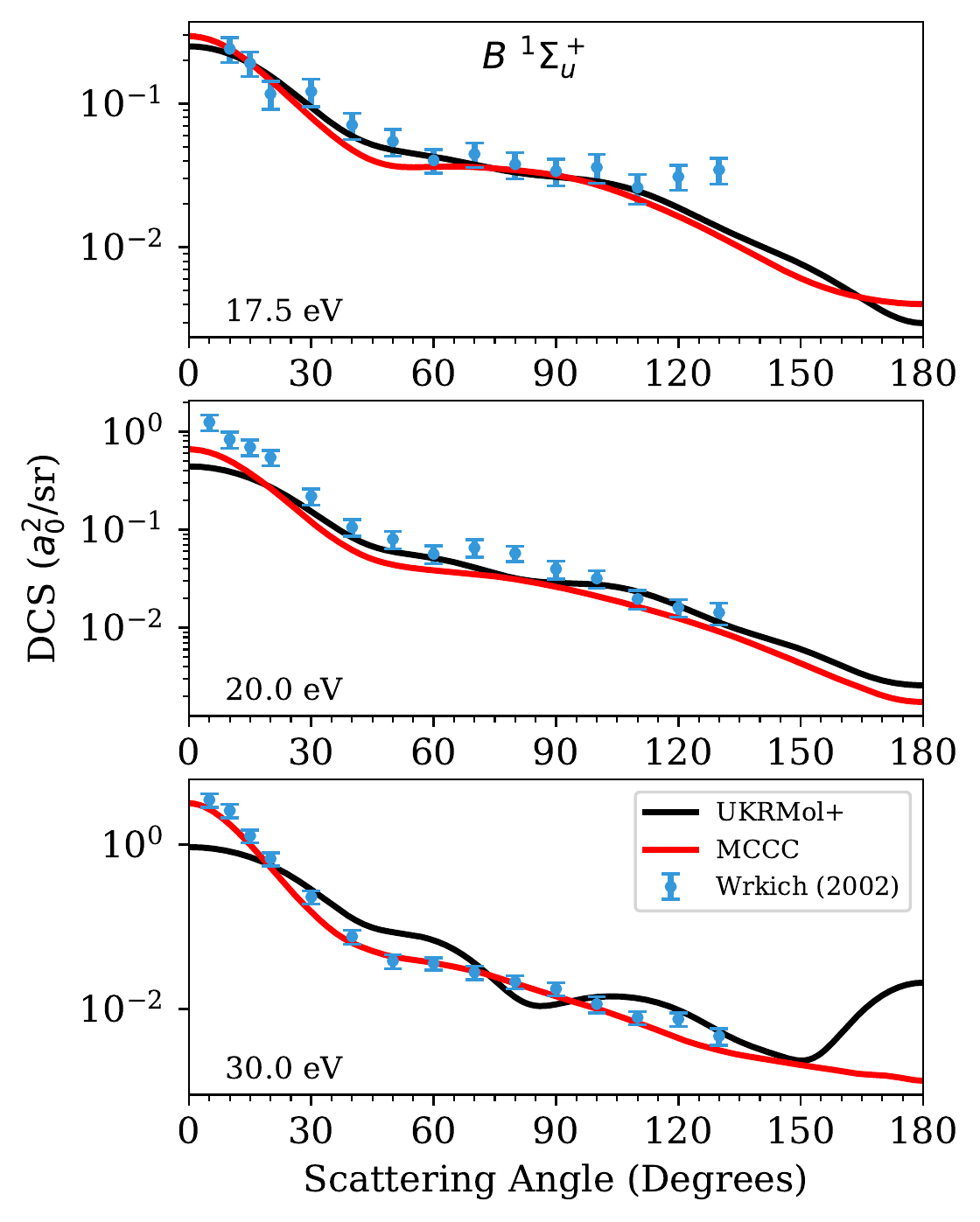}
  \caption{ DCS for the $X \ ^1\Sigma_g^+ \rightarrow B \ ^1\Sigma_u^+ $ transition. Comparison of
  the UKRMol+ and MCCC calculations with the measurements of \citeasnoun{Wrkich2002}.}
  \label{fig:dcs_state3}
\end{figure}

\begin{figure}[htbp]
  \centering
  \includegraphics[width=1.0\linewidth]{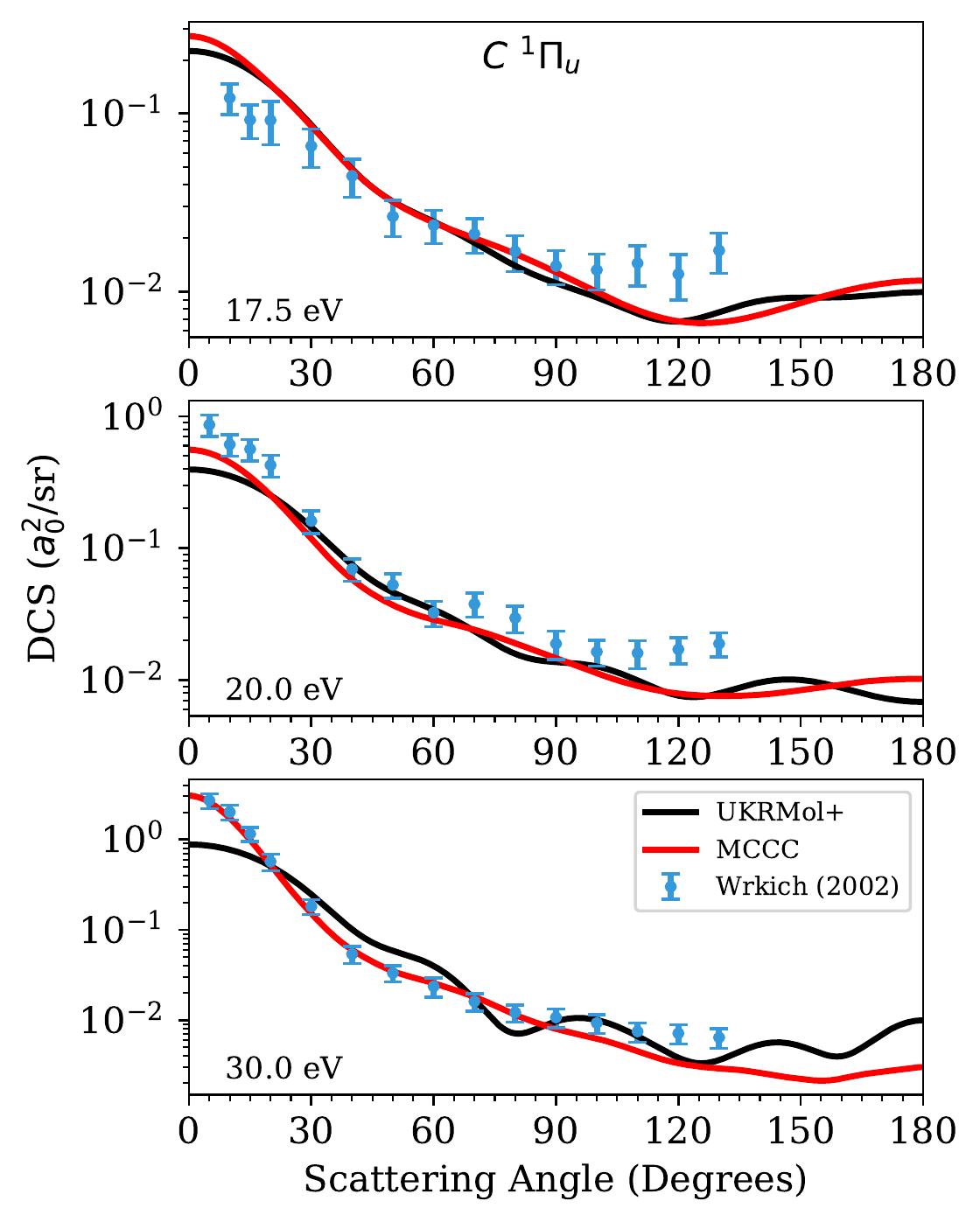}
  \caption{ DCS for the $X \ ^1\Sigma_g^+ \rightarrow C \ ^1\Pi_u $ transition. Comparison of the
  UKRMol+ and MCCC calculations with the measurements of \citeasnoun{Wrkich2002}.}
  \label{fig:dcs_state6}
\end{figure}

DCSs could not be determined in the emission experiments of \citeasnoun{Liu1998}. However,
\citeasnoun{Wrkich2002} produced a set of EELS DCS which have been plotted in Figs.
\ref{fig:dcs_state3} and \ref{fig:dcs_state6}. For the $B \ ^1\Sigma_u^+$ state the agreement
with experiment is good. At 30 eV however the R-matrix calculation displays oscillations that are
not present in the MCCC Calculation. This is due to a lack of convergence in the partial-wave
expansion. Typically a Born correction would be applied to dipole allowed transitions. However, in
the present work this has not been possible. The Born top-up requires a sufficiently converged
cross-section, up to some intermediate number of partial waves, $\bar{L}_{\mathrm{max}}$. For MCCC
this is found to be $\bar{L}_{\mathrm{max}}=25$, or more, depending on the scattering energy
\cite{Zammit2017a}. A similar approach was attempted for the R-matrix calculation. Although this was
not tractable given currently available software and computational power (see
\ref{sec:top_up_methods}).

\begin{figure}[htbp]
  \centering
  \includegraphics[width=1.0\linewidth]{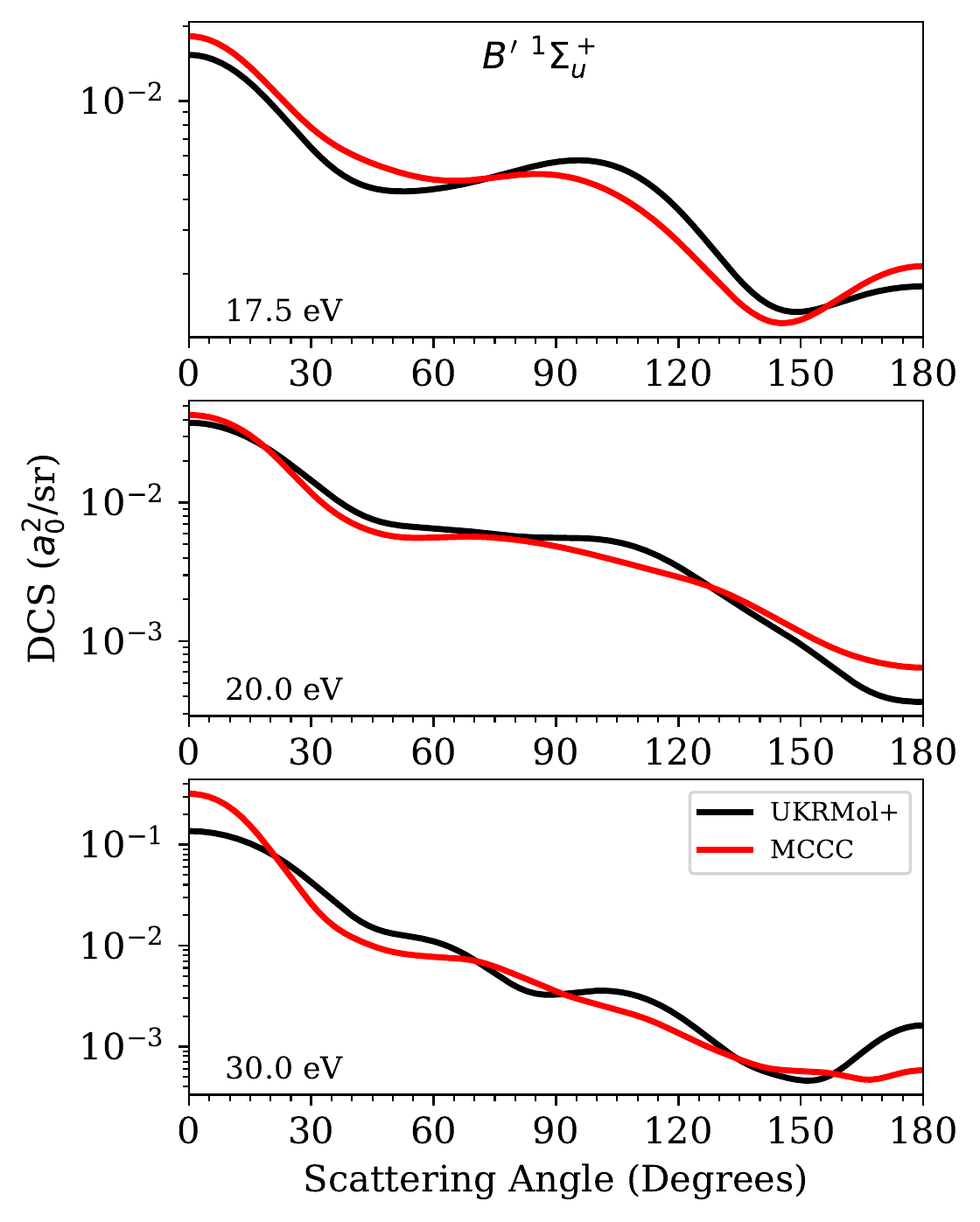}
  \caption{ DCS for the $X \ ^1\Sigma_g^+ \rightarrow B^\prime \ ^1\Sigma_u^+ $ transition.
  Comparison of the UKRMol+ and MCCC calculations.}
  \label{fig:dcs_state9}
\end{figure}

\begin{figure}[htbp]
  \centering
  \includegraphics[width=1.0\linewidth]{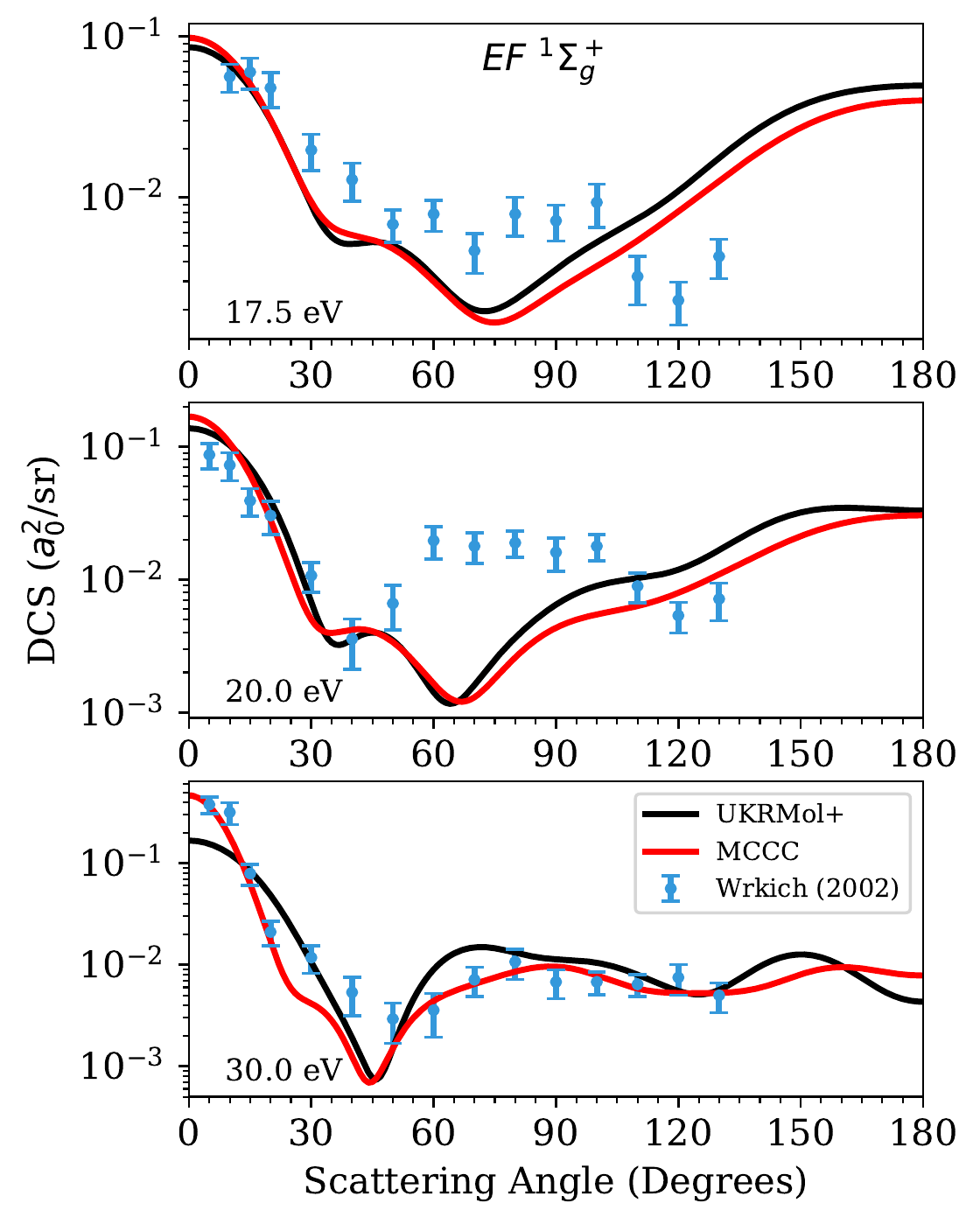}
  \caption{ DCS for the $X \ ^1\Sigma_g^+ \rightarrow EF \ ^1\Sigma_g^+ $ transition. Comparison of
  the UKRMol+ and MCCC calculations with the measurements of \citeasnoun{Wrkich2002}.}
  \label{fig:dcs_state5}
\end{figure}

Similarly, the oscillations observed for the $B \ ^1\Sigma_u^+$ state at higher energies are also
observed in states $C \ ^1\Pi_u $ and $B^\prime \ ^1\Sigma_u^+$ (Figs. \ref{fig:dcs_state6} and
\ref{fig:dcs_state9}). Furthermore, in all of the singlet state DCSs, Figs. \ref{fig:dcs_state3},
\ref{fig:dcs_state6}, \ref{fig:dcs_state9} and \ref{fig:dcs_state5}, the R-matrix calculation has a
lower forward peak. This is attributed, as in the elastic scattering case, to a lack of convergence
in the number of partial waves used. Regardless, forward and backward scattering only make a small
contribution to the total ICS. Therefore the differences caused by the oscillatory behaviour and
lower forward peak are lost upon integration. This highlights the importance of using DCSs as a
stringent test of theories. Two theories may produce the same ICS but have different angular
profiles.

In contrast to the dipole-allowed singlet states, the forbidden $EF \ ^1\Sigma_g^+$ state DCS (Fig.
\ref{fig:dcs_state5}) is not as sensitive to higher partial-waves. Agreement between the two
theories is good. The agreement between theory and experiment is acceptable, except for the
scattering angles from 60\dgr\ to 100\dgr\ at 20 eV where the experiment gives a larger
cross-section, which could be due to the analysis of the measured EELS.

\begin{figure}[htbp]
  \centering
  \includegraphics[width=1.0\linewidth]{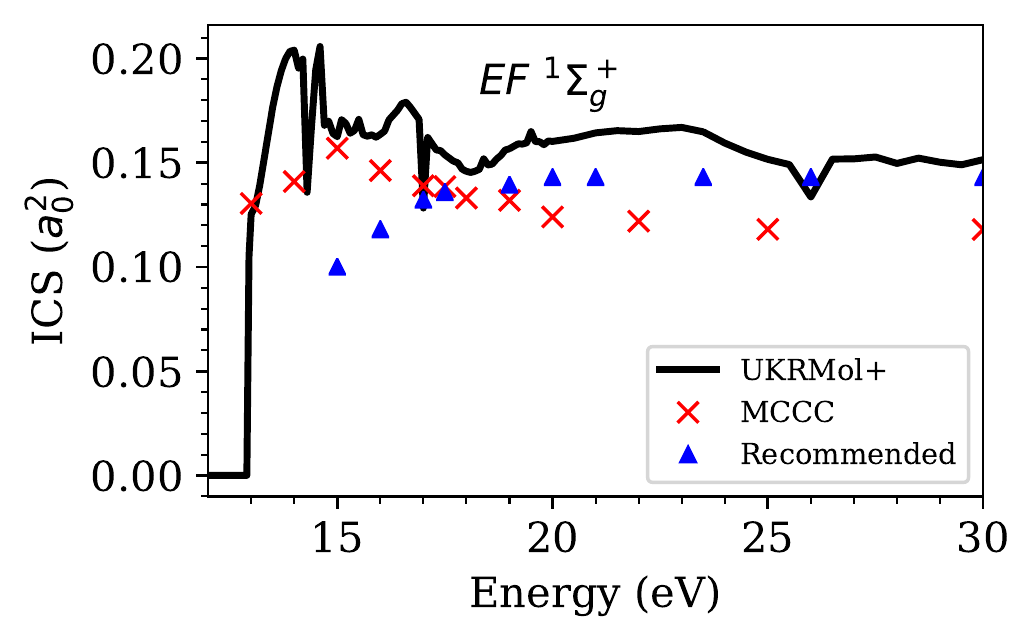}
  \caption{ ICS for the $X \ ^1\Sigma_g^+ \rightarrow EF \ ^1\Sigma_g^+ $ transition. Comparison of
  the UKRMol+ and MCCC calculations with the recommended data of \citeasnoun{Yoon2008}.}
  \label{fig:ics_state5}
\end{figure}

Comparing the ICS (Fig. \ref{fig:ics_state5}) between the two theories, the R-matrix calculation is
consistently above the MCCC data. Again, this is due to the absence of ionisation channels in the
R-matrix close-coupling expansion that leads to an overestimated cross-section.

The recommended data are based on an emission experiment carried out by \citeasnoun{Liu2003}. Whilst
the $EF \ ^1\Sigma_g^+$ state is dipole-forbidden, the cross-section can be inferred using a
combination of theoretical and experimental considerations. There is a difference in threshold for
experiment, which occurs near 15 eV as opposed to 13 eV for the FN MCCC and R-matrix calculations.
However the magnitude and qualitative shape agree with theory.

\subsection{Adiabatic-Nuclei Cross-Sections}
\label{sub:adiabatic_nuclei}

\begin{figure}[htpb]
  \centering
  \includegraphics[width=1.0\linewidth]{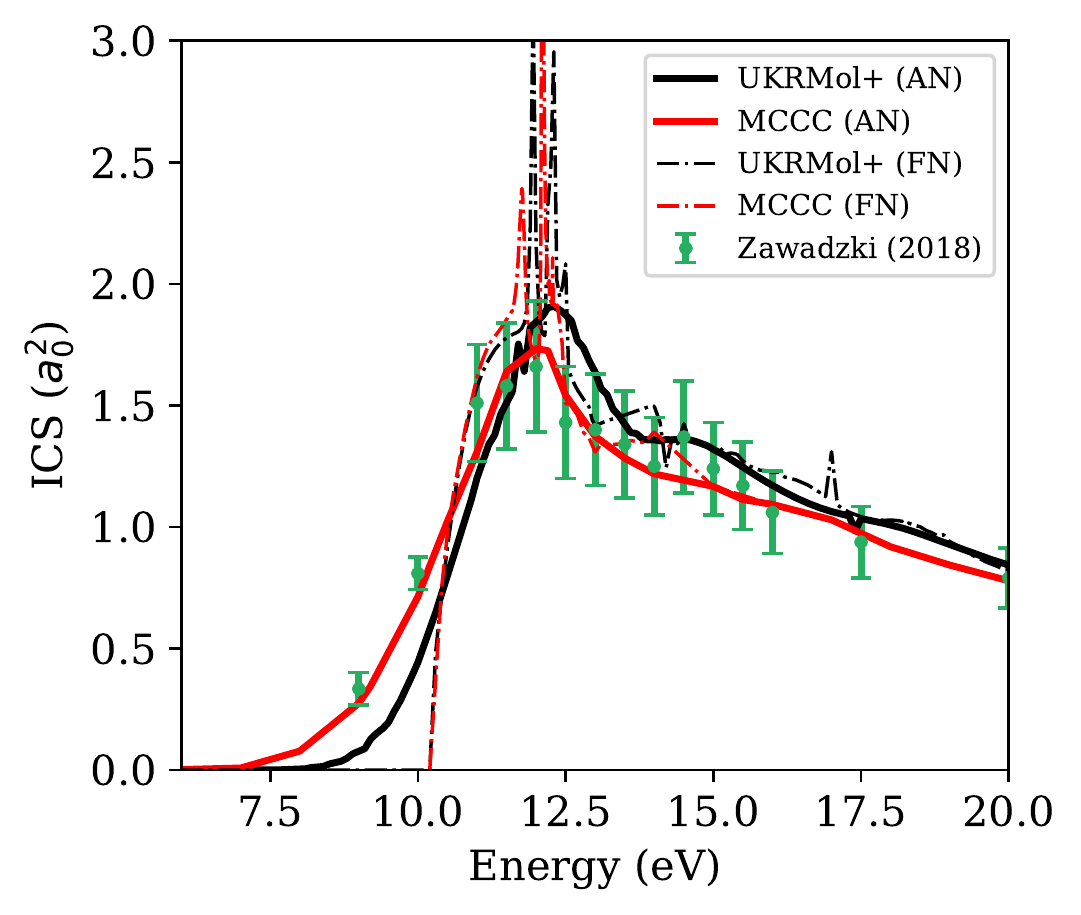}
  \caption{ Integrated cross-section for the $X \ ^1\Sigma_g^+ \rightarrow b \ ^3\Sigma_u^+ $
    transition using the AN approximation.  (black) UKRMol+, (red) MCCC, (green dots) experiment
    from \citeasnoun{Zawadzki2018}.
  }
  \label{fig:ANapproximation}
\end{figure}

In this section we make use of the AN approximation described previously. In Fig.
\ref{fig:ANapproximation} both FN (dot-dashed line) and AN calculations (solid line) are shown
side-by-side for electronic excitation to the first excited state ($X \ ^1\Sigma_g^+ \rightarrow b \
^3\Sigma_u^+ $). For both the MCCC (red) and UKRMol+ (black) calculations we can see two main
differences. The first is that resonant structures are washed-out and the second is that the sharp
turn-on near the vertical excitation threshold (10 eV) is smoothed into a ramp. This is due to the
vibrational averaging over different molecular geometries. The threshold for the $X \ ^1\Sigma_g^+
\rightarrow b \ ^3\Sigma_u^+ $ transition is essentially the vertical excitation energy. For some
geometries this will be lower than 10 eV and for others will be greater. The average is weighted by
the square of the ground vibrational wavefunction, which means the largest contributions occur at
the maximum of the wavefunction i.e., about R$_0$. This is why the FN calculation at $R=R_0$ and the
AN calculation are broadly similar. Adiabatic effects have consequences for near-threshold electron
impact dissociation of H$_2$ \cite{jt229}.

The AN approximation requires FN calculations to be performed across a grid of different
internuclear bond separations. For the R-matrix calculations, a grid size of $\Delta R = 0.05$ a.u.
was used for $0.95 < R < 1.95$ a.u., with a finer grid of $\Delta R = 0.01$ a.u. used in the region
closer to the mean vibrational bond length, $1.35 < R < 1.55$ a.u. Due to the large number of FN
calculations required it was not possible to use the full model described previously. Therefore a
smaller model was used which featured a singly augmented aug-cc-pVTZ basis set and an R-matrix
radius $a=25$ a.u. The smaller radius allowed the continuum representation to be simplified to 22
BTOs per angular momentum symmetry with $L_\mathrm{max} = 4$ without sacrificing completeness. As
before, all of the target states below 30 eV were included which led to a 59-state model. This model
works well for the first excited state but due to the simplified target description it cannot
represent higher-excited states.

\section{Conclusion}%
\label{sec:conclusion}

In this paper we demonstrate good agreement with recent experimental data
\cite{Muse2008,Zawadzki2018}, validated by two independent theories. The agreement with the
recommended ICS data \cite{Yoon2008} and older experimental data \cite{Wrkich2002} is worse,
predominantly for the triplet states but we believe this is due to the difficulties associated with
the underlying experiments. That is, it is difficult for experiments to separate the overlapping
contributions coming from different triplet excited states and therefore the error margin is larger
for these types of experiment. Any other significant differences between the two theories and
experiments are well understood.

This is the first time the CCC and R-matrix theories have been verified for a molecular target. This
work presents one of the largest molecular R-matrix calculations to date. Many novel features have
been exploited for the first time: a triply-augmented target basis set, a box size of 100 a.u. and
the first B-spline only continuum for a molecular target. This shows that both MCCC and R-matrix
method can be used to perform large-scale, high-accuracy close-coupling calculations.

We have compared both fixed-nuclei and adiabatic-nuclei cross-sections obtained using the R-matrix
and MCCC methods. For FN calculations, dipole-forbidden states generally show better agreement in
the DCSs. Dipole-forbidden states do not require a born top-up and generally converge quicker for
the same number of partial waves, compared to dipole allowed transitions \cite{Zammit2017a}. For the
dipole-allowed states the R-matrix calculations show oscillatory behaviour but this could be
eliminated by using a higher cutoff in the number of partial-waves. However, this is currently not
tractable given currently available hardware and software.

All of the ICSs show good agreement between the two theories with the exclusion of weak transitions
that are more sensitive to the absence of ionisation channels in the R-matrix calculations, leading
to slightly enhanced cross-sections. The AN ICS for the first excited state shows excellent
agreement between the two theories and the recent experimental data.

There are several directions for future work. Firstly, it would be interesting to compare the effect
of target model used in the MCCC calculations i.e., spherical versus spheroidal. Preliminary results
for the $EF \ ^1\Sigma_g^+$ state suggest that the use of a spheroidal model could improve the
agreement between both theories.

Secondly, in order to accurately describe ionisation effects in the R-matrix method we would need to
employ the RMPS method. Whilst the RMPS method is implemented in UKRMol+ the calculations for this
system are currently too expensive.

Additionally, for the R-matrix calculations presented in this work we have not been able to carry
out systematic, quantitative analysis of the uncertainties. This is a common problem across the
field for theoretical calculations \cite{jt642}. For future work, we seek a tractable approach that
is capable of providing uncertainties for our calculated data.

Finally, a general approach for handling Born top-ups, similar to the ABS method used in MCCC
calculations would be desirable for the UKRMol+ calculations in order to reach convergence where
larger numbers of partial-waves are required (as discussed in \ref{sec:top_up_methods}).

\section*{Acknowledgements}

TM is funded by EPSRC (grant No. EP--M507970--1). The R-matrix calculations were carried out using
UCL Myriad computing resources. ZM acknowledges support of the PRIMUS project (No. 116-45/247084) of
Charles University. The MCCC calculations were carried out with support from the United States Air
Force Office of Scientific Research, Curtin University, Los Alamos National Laboratory (LANL), and
resources provided by the Pawsey Supercomputing Centre, with funding from the Australian Government
and Government of Western Australia. LHS acknowledges the contribution of an Australian Government
Research Training Program Scholarship, and the support of the Forrest Research Foundation. MCZ
would like to specifically acknowledge LANL's ASC PEM Atomic Physics Project for its support. LANL
is operated by Triad National Security, LLC, for the National Nuclear Security Administration of the
U.S. Department of Energy under Contract No. 89233218NCA000001.

\section*{References}

\appendix
\section{Including Higher Partial Waves}
\label{sec:top_up_methods}

To include higher partial waves, specifically for dipole-allowed transitions, we require a top-up
procedure. In R-matrix calculations this is done using an approach suggested by
\citeasnoun{Norcross1982}. The MCCC uses an equivalent method described in \cite{Zammit2017}. For a
DCS the top up procedure is given by
\begin{eqnarray}
\frac{d\sigma}{d\Omega} =&& \left(\frac{d\sigma}{d\Omega}\right)_{Born} + \nonumber\\
&&\sum_{\lambda=0}^{\lambda_{max}} (A_{\lambda}-A_{\lambda}^{Born})P_{\lambda}(\cos\theta),
\end{eqnarray}
where the first term on the right hand side is the DCS calculated for inelastic dipolar scattering
in the first Born approximation and the second term includes the contribution of the lower partial
waves $A_{\lambda}$ calculated with close-coupling and subtraction of the corresponding Born partial
waves $A_{\lambda}^{Born}$. Only orientational averaging of the molecule is taken into account. This
approach was used in previous R-matrix calculations for inelastic collisions, but only for ICSs
\cite{jt256,jt429,Masin2012}, which tend to converge quicker than DCSs. Recently, \citeasnoun{jtCO}
employed the Born correction described above for the electronically inelastic DCS of CO. This
method, however, requires a sufficiently high partial wave cutoff, $L_\mathrm{max}$. At lower
partial waves the analytic Born method is less accurate and tends to overestimate the cross-section,
leading to unphysical negative cross-sections.

Born corrections have been successfully applied to DCSs for elastic collisions, see e.g.,
\citeasnoun{jt464} and \citeasnoun{Masin2012}. However, these cross-sections are usually an order of
magnitude larger than those for dipole-allowed inelastic transitions. Hence, they are less
susceptible to the oscillatory behaviour seen in inelastic DCSs.

\begin{figure}[htpb]
  \centering
  \includegraphics[width=1.0\linewidth]{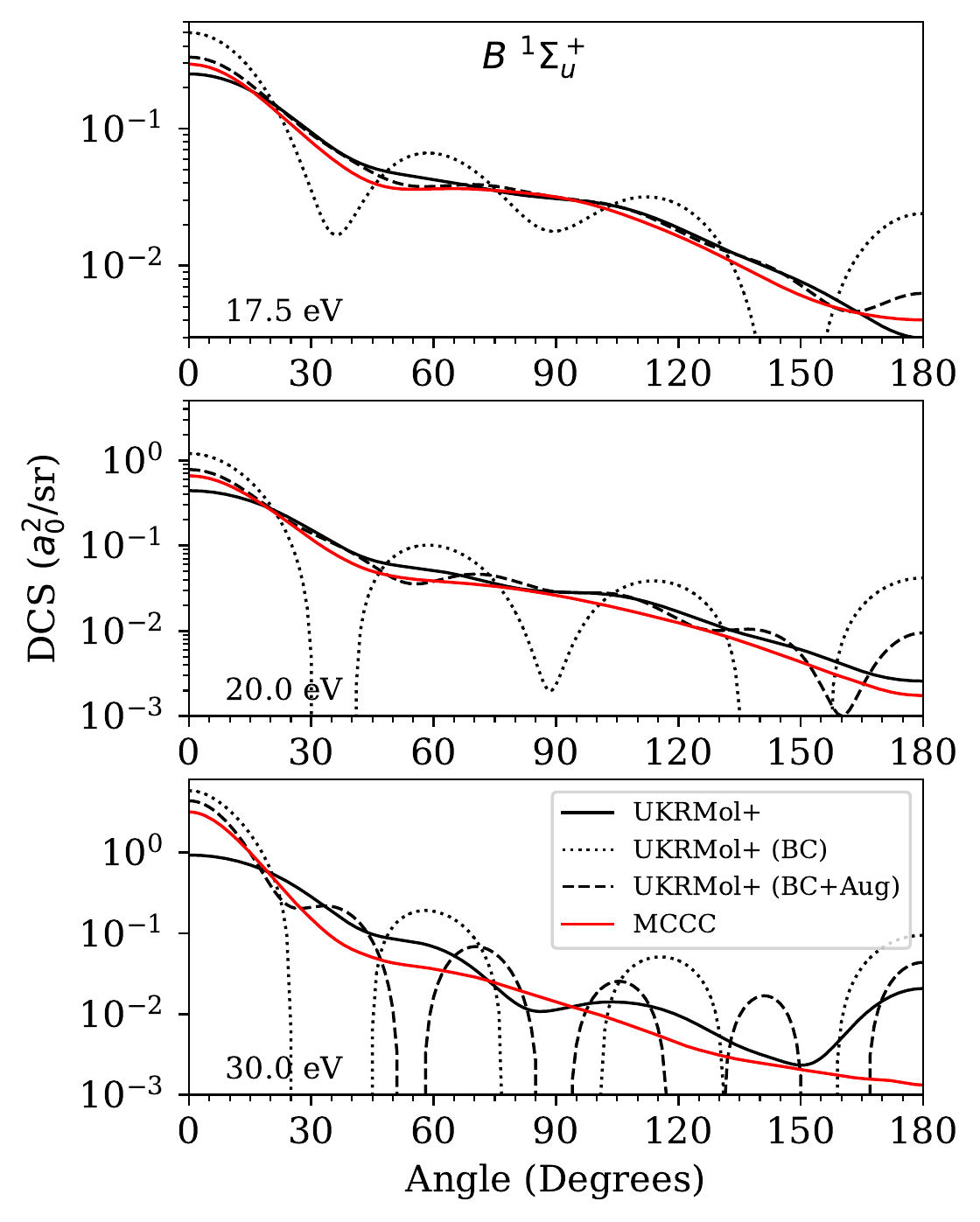}
  \caption{ DCS for the $B \ ^1\Sigma_u^+$ state calculated using the R-matrix method, with
  $L_\mathrm{max}=6$ (black) and the MCCC method (red). The Born top-up was applied to the original
calculation before (dotted) and after (dashed) the T-matrix elements were augmented with a cheaper
$L_\mathrm{max}=10$ calculation.}
  \label{fig:topup}
\end{figure}

Figure \ref{fig:topup} shows the DCS for the dipole-allowed $B \ ^1\Sigma_u^+$ state. In solid black
we have the original R-matrix calculation without the Born correction. If we apply the Born
correction to the DCS we obtain the dotted line. At 17.5 eV, the Born corrected DCS displays
unphysical behaviour around 150\dgr\ where it becomes negative. The situation worsens for higher
energies. This is due to an incomplete convergence of the partial-wave Born contribution
$\sum_{\lambda} A_{\lambda}^{Born}P_{\lambda}(\cos\theta)$.

To resolve this issue, the MCCC approach \cite{Zammit2017a} has been to run a smaller-sized
calculation but with a higher cutoff e.g., $\bar{L}_\mathrm{max}=25$. The results of this
calculation are then used to augment the T-matrices of the more expensive calculation. This allows
the DCS contributions from higher partial waves to be calculated with the more accurate MCCC theory
before including the additional contributions from the Born procedure.

A similar approach has been adopted in the R-matrix calculations, however $\bar{L}_\mathrm{max}=25$
is currently not computationally feasible with the UKRMol+ codes. Calculations using a smaller
model, but with $\bar{L}_\mathrm{max}=10$, have been computed and these were used to augment the
T-matrices of the accurate R-matrix calculation with $L_\mathrm{max}=6$. When augmenting the
T-matrix elements, care must be taken to phase-match the two calculations. This can be achieved by
comparing the transition dipole moments of the target states involved in each transition.

The result of augmenting the T-matrices and applying the Born correction is shown as the dashed line
in Fig. \ref{fig:topup}. For the lowest scattering energy shown, 17.5 eV, the oscillatory behaviour
is greatly reduced and the Born correction improves the quality of agreement between the MCCC and
R-matrix calculations. At 20 eV Born-corrected DCS is improved but it still shows oscillatory
behaviour that is characteristic of a lack of convergence. At 30 eV, even with the augmented
T-matrix elements the DCS remains oscillatory when the Born correction is applied.

In theory, an approach similar to the MCCC method can be developed for the R-matrix calculations but
there are two factors that currently inhibit further improvement. The first is that the target
states from cheaper calculations need to be shifted to the more accurate values from the expensive
calculation. For the R-matrix calculations, presented in this work, the energies were shifted in the
outer-region. This is not ideal and instead we need to implement the energy shift in the $N+1$
scattering calculation, similar to the approach used by \citeasnoun{jt208}. Secondly, the
outer-region quickly dominates the computational resources required, both physical RAM and CPU-time,
as a large number of channels are generated for higher partial waves. Furthermore a sophisticated
approach would need to be implemented in the outer-region to reduce the number of states included in
the calculation.

As an alternative approach, we also attempted to top-up the DCS using a more basic method (not
shown). We ran two cheaper calculations with small basis sets using $L_\mathrm{max}=6$ and
$L_\mathrm{max}=10$. We took the difference between the two DCSs and used this to top-up the
expensive calculation. This approach does help to capture the forward peak scattering but it was too
susceptible to unphysical negative cross-sections when the differences between the cheap
calculations became negative. This method behaved particularly poorly in regions where the
cross-section was small.

In summary, we believe the MCCC approach to the Born top-up is the most sensible way forward,
however there is still work to be done before it can be implemented in R-matrix calculations.

%

\end{document}